\documentclass[sigconf]{acmart}

\AtBeginDocument{%
  }

\setcopyright{rightsretained}
\copyrightyear{2024}
\acmYear{2024}
\acmDOI{10.1145/3658644.3690239}
\acmConference[CCS '24]{Proceedings of the 2024 ACM SIGSAC Conference on Computer and Communications Security}{October 14--18, 2024}{Salt Lake City, UT, USA}
\acmISBN{979-8-4007-0636-3/24/10}
\acmBooktitle{Proceedings of the 2024 ACM SIGSAC Conference on Computer and Communications Security (CCS '24), October 14--18, 2024, Salt Lake City, UT, USA}

\usepackage[nolist]{acronym}
\usepackage{xspace}

\usepackage{tabularx}
\usepackage{longtable} 
\usepackage{colortbl}
\usepackage{hhline}
\usepackage{graphicx} 
\usepackage{pifont} 
\usepackage{multirow} 

\usepackage{caption}

\usepackage{listings}
\usepackage{color}

\usepackage{subcaption}
\usepackage{adjustbox}
\usepackage{changepage}

\definecolor{mygreen}{rgb}{0,0.6,0}
\definecolor{mygray}{rgb}{0.5,0.5,0.5}
\definecolor{mymauve}{rgb}{0.58,0,0.82}

\def\CorrEntropy{Correlation Entropy}
\def\AbsEntropy{Absolute Entropy}

\newcommand{\mypar}[1]{\smallskip\noindent\textbf{#1.}\xspace}

\begin{document}

\begin{acronym}
	\acro{ASLR}[ASLR]{Address Space Layout Randomization}
  \acro{OS}[OS]{Operating System}
  \acroplural{OS}[OSes]{Operating Systems}
  \acro{VM}[VM]{Virtual Machine}
  \acroplural{VM}[VMs]{Virtual Machine}
  \acro{TLS}[TLS]{Thread Local Storage}
  \acroplural{TLS}[TLSes]{Thread Local Storages}
\end{acronym}

\title{The Illusion of Randomness: An Empirical Analysis of Address Space Layout Randomization Implementations}



\author{Lorenzo Binosi}
\authornote{Both authors contributed equally to this research.}
\email{lorenzo.binosi@polimi.it}
\affiliation{%
  \institution{Politecnico di Milano}
  \city{Milan}
  \country{Italy}
}

\orcid{1234-5678-9012}
\author{Gregorio Barzasi}
\authornotemark[1]
\email{gregorio.barzasi@mail.polimi.it}
\affiliation{%
  \institution{Politecnico di Milano}
  \city{Milan}
  \country{Italy}
}

\author{Michele Carminati}
\email{michele.carminati@polimi.it}
\affiliation{%
  \institution{Politecnico di Milano}
  \city{Milan}
  \country{Italy}
}

\author{Stefano Zanero}
\email{stefano.zanero@polimi.it}
\affiliation{%
  \institution{Politecnico di Milano}
  \city{Milan}
  \country{Italy}
}

\author{Mario Polino}
\email{mario.polino@polimi.it}
\affiliation{%
  \institution{Politecnico di Milano}
  \city{Milan}
  \country{Italy}
}

\renewcommand{\shortauthors}{Binosi et al.}

\begin{abstract}
  \ac{ASLR} is a crucial defense mechanism employed by modern operating systems to mitigate exploitation by randomizing processes' memory layouts. However, the stark reality is that real-world implementations of \ac{ASLR} are imperfect and subject to weaknesses that attackers can exploit. This work evaluates the effectiveness of \ac{ASLR} on major desktop platforms, including Linux, MacOS, and Windows, by examining the variability in the placement of memory objects across various processes, threads, and system restarts. In particular, we collect samples of memory object locations, conduct statistical analyses to measure the randomness of these placements and examine the memory layout to find any patterns among objects that could decrease this randomness. The results show that while some systems, like Linux distributions, provide robust randomization, others, like Windows and MacOS, often fail to adequately randomize key areas like executable code and libraries. Moreover, we find a significant entropy reduction in the entropy of libraries after the Linux 5.18 version and identify correlation paths that an attacker could leverage to reduce exploitation complexity significantly. Ultimately, we rank the identified weaknesses based on severity and validate our entropy estimates with a proof-of-concept attack. In brief, this paper provides the first comprehensive evaluation of \ac{ASLR} effectiveness across different operating systems and highlights opportunities for \ac{OS} vendors to strengthen \ac{ASLR} implementations.
\end{abstract}



\received{20 February 2007}
\received[revised]{12 March 2009}
\received[accepted]{5 June 2009}

\maketitle

\section{Introduction}
\label{sec:introduction}

In the context of software exploitation, the more information an attacker can gather about the targeted system, the easier it is to exploit it. In particular, one crucial piece of information for the success of \textit{memory errors attacks} (e.g., buffer overflows) is the exact address of objects inside the memory space of the running program; this information allows an attacker to successfully exploit the software vulnerabilities hijacking the execution flow of the program or leaking sensitive data. Even if these vulnerabilities are well known, they still account for more than 20\% of the CVE reported \cite{CVE_stats}.

\textit{Address Space Layout Randomization (ASLR)} is a security measure developed to improve resilience against exploitation techniques that depend on precise memory locations. The base mechanism of \ac{ASLR} is the randomization of the memory layout of processes to make exploitation a \textit{game of chance}. Thus, its effectiveness increases as the number of attempts needed to hit a precise object in memory increases. Ideally, the entropy of the memory layout should be high enough to make the chances of a successful attack negligible; we see that this is not always the case in real-world implementations.
Moreover, we would like \ac{ASLR} to randomize every object that is allocated inside the memory of a program at the moment of allocation and with high entropy (so, low predictability on the object position); unfortunately, there is a gap between theoretical capabilities and real-world implementation: as shown in this work, on many systems, \ac{ASLR} lacks at least one of the mentioned aspects. In particular, the memory is often grouped in sections or groups of sections and then randomized all together at the program launch; as a consequence, even if the entropy of single memory objects is high, it can be reduced by collecting information about other memory objects, as their memory sections of origin may be correlated~\cite{Gisbert2016ExploitingLA_PositiveEnt}. 
Even though the correlation is an already known vulnerability~\cite{DBLP:conf/uss/HerlandsHD14}, current research still lacks a comprehensive evaluation regarding the severity of this vulnerability in terms of entropy reduction. Moreover, the scope of these studies is often limited, focusing on specific sections and predominantly considering operating systems tailored towards Linux enterprise solutions~\cite{marco2019address}. Additionally, the recent shift to ARM architecture in Apple devices has not been sufficiently addressed, leaving the effectiveness of ASLR on such systems unknown. In this work, we directly address these gaps. 

In summary, our contributions are the following:

\begin{itemize}

    \item We conduct an empirical evaluation of the \ac{ASLR} mechanism's effectiveness across various operating systems, including Linux, MacOS, and Windows, focusing on allocation probability and the impact of this probability on the correlation between memory objects.
    
    \item We identify instances where the absolute entropy of relevant memory objects is considerably low, which makes bruteforce attacks feasible. Moreover, even when the absolute entropy is high, we identified high correlation paths that can be exploited to reduce guessing space and speed up bruteforce exploits. 
    
    \item We identify memory objects with non-uniform distributions, meaning that certain addresses are more likely to be chosen for allocating a memory object, consequently reducing its overall entropy.
  
    \item We identify a sudden reduction in randomization entropy of shared libraries in Linux systems since the 5.18 kernel release due to the introduction of Linux Folios performance optimization.

    \item We demonstrate through empirical proofs that current randomness in \ac{ASLR} implementations can be easily bypassed in a short amount of time, even on the latest versions of the analyzed operating systems. Consequently, we discuss possible mitigation strategies to improve the effectiveness of ASLR.
    
    \item We release the code used to sample all memory objects in the \acp{OS} under study, as well as the code for our \ac{ASLR} analysis. Additionally, we provide the dataset employed in our analysis, which can be used to replicate our results or conduct further research. All the materials can be found at: \url{https://zenodo.org/doi/10.5281/zenodo.12786166}.
    
\end{itemize}

\section{Background and Motivation}
\label{sec:motivations}

Analyzing \ac{ASLR} performance is crucial, as \ac{ASLR} represents one of the final security defenses that need to be overcome for exploiting a system. It is worth noting that \ac{ASLR} is not the sole defense mechanism against attacks; indeed, additional protections such as the \textit{NX} bit and \textit{Stack Smashing Protection (SSP)} are also in place to enhance security in case a new vulnerability emerges. However, the fact that we can rely on other security mitigations does not excuse vendors from implementing a robust \ac{ASLR} mechanism. Moreover, being aware of the specific \acp{OS}' limitations is particularly useful when choosing which best fits a particular scenario's security requirements and threat model. 

In the subsequent sections, we examine the most common \ac{ASLR} weaknesses, justify the selection of the specific operating system under analysis in this work, and provide a threat model where \ac{ASLR} should mitigate possible attacks.

\subsection{Evaluate ASLR Performances}
\label{subsec:metrics}

To understand the limitations of \ac{ASLR} implementation, we start by defining its goals and expected performance: to increase the average number of attempts an attacker must make to correctly guess a memory address through the obfuscation of the memory layout. In other words, it is a mechanism that frequently randomizes memory objects with low predictability and high granularity. More formally, we utilize the taxonomy proposed by Marco-Gisbert and Ripoll~\cite{marco2019address}, which classifies the capabilities of \ac{ASLR} implementations into three categories: when, what, and how. By examining the performance in these categories, we assess the effectiveness of an \ac{ASLR} implementation and identify areas for improvement.

\mypar{When} The differentiation in \ac{ASLR} mechanisms across various \acp{OS} lies when they perform randomization, i.e., in the frequency with which they refresh the randomization of their memory sections. Ideally, a system would randomize every memory object at the moment of allocation; however, this is challenging due to performance degradation and implementation complexity. Consequently, no commercial \ac{OS} currently adopts such an advanced technique. The most practical goal is to have the memory randomized every time a new process is allocated in memory (as in the case of Linux). In contrast, some \acp{OS} like MacOS~\cite{marco2019address} and Windows~\cite{aristizabal2013measuring,DBLP:conf/acsac/LiJS06} only randomize parts of memory and only at boot. However, this approach is considered a flawed implementation of \ac{ASLR}, as a local attacker could use information gathered from other processes to locate the library position precisely; moreover, a remote attacker could bruteforce the address starting from the lowest possible value, to the highest one, or perform a byte-per-byte attack~\cite{byte-per-byte}.

\mypar{What} Another critical aspect concerns the granularity of what is randomized and the number of sections randomized: even a single non-randomized object can be leveraged to take over a system. For instance, certain implementations fail to randomize all executable objects or do so only at boot-time, significantly reducing the system's overall security and increasing the likelihood of successful Return-to-Library (Ret2Lib) or Return-Oriented Programming (ROP) attacks.

\mypar{How} The last category concerns how objects are randomized, specifically in terms of the number of bits randomized and the relative positioning of the objects. Current implementations of \ac{ASLR} primarily adopt the \emph{Partial-VM} randomization strategy. This method partitions virtual memory into segments where objects are placed. For instance, on a 64-bit Windows 11 OS, libraries and the executable are allocated in a specific region ranging from the address \texttt{0x7ff800000000} to \texttt{0x800000000000}; no other object can be allocated in such a region. Although this approach significantly simplifies the \ac{ASLR} implementation, it also increases the likelihood of generating strong correlation paths.

\subsection{ASLR Implementation Weaknesses}
\label{subsec:weakness}

All aspects mentioned in Section~\ref{subsec:metrics} impact the unpredictability of memory layout, which can be quantified using entropy. In particular, we refer to \AbsEntropy\ as the number of bits that represent the randomization of a memory object. If the probability of allocating an object follows a uniform distribution, then the \AbsEntropy\ would be the base-2 logarithm of the total number of distinct positions a memory object can occupy within the virtual address space. Nonetheless, given that real-world implementations might not follow a uniform distribution, we consider the more general case following the Shannon Entropy formula: $H = -\sum_{i=1}^{n} p(x_i) \log_2 p(x_i)$, where \textit{n} is the total number of possible distinct positions the memory object can occupy, and \textit{$p(x_i)$} is the probability of the memory object occupying the i-th position. We consider \textit{bits} as our reference measure since, under a uniform distribution, the entropy value is the number of bits altered in an address by ASLR.
Instead, we refer to \CorrEntropy\ as the number of bits representing the randomization between two memory objects, for which we also utilize the Shannon Entropy formula. Here, \textit{n} denotes the total number of potential differences between two memory objects in terms of memory addresses. 

The lack of \AbsEntropy\ and \CorrEntropy\ is the primary weakness of ASRL implementations. 

\subsubsection{Low \AbsEntropy}
\label{subsubsec:weakness_low_entropy}

The major problem affecting \ac{ASLR} implementations is low \AbsEntropy. In this context, low \AbsEntropy\ is directly associated with the amount of bruteforce effort an attacker needs to perform to correctly guess the position of a memory object without using any particular technique to predict the position. 

In modern \acp{OS}, the \texttt{MSb} divides the user space virtual memory (\texttt{MSb} = 0) from the kernel space virtual memory (\texttt{MSb} = 1), and thus, only 47 bits are available for addressing memory in user space. Therefore, due to the page offset, the maximum entropy achievable in a 64-bit system is $47 bits - 12 bits = 35 bits$ for an Intel x86\_64 system and $47 bits - 14 bits = 33 bits$ for an ARM M1 system. Unfortunately, as empirically shown in this paper, no 64-bit OS reaches these entropy measures, as additional factors influence its estimation. Firstly, certain sections, such as the heap and stack, require room to expand, limiting the potential for placing such objects in memory since they necessitate space to grow upwards or downwards. Secondly, non-contiguous sections like the heap and libraries must have large memory ranges available for random placement. Thirdly, flawed RNG mechanisms can also contribute to reduced entropy.

\subsubsection{Low \CorrEntropy}
\label{subsubsec:weakness_low_corr_entropy}

The correlation between various memory objects can significantly impact the complexity of an attack, as the leak of one memory address may significantly reduce the effort needed to compromise another. This reduction in complexity can be quantified by assessing \CorrEntropy, which measures the entropy of the offset between two memory objects. We distinguish two scenarios:

\mypar{Correlated Objects} When the addresses of allocated memory objects are correlated, their \CorrEntropy\ is lower than the \AbsEntropy\ of either object individually. Consequently, a vulnerability that leaks the address of one object can be exploited to de-randomize other objects. In other words, the knowledge of the address of one object reduces the search space for the address of the other one.

\mypar{Independent Objects} Two randomly allocated memory objects are independent when the \CorrEntropy\ between these objects is higher or equal than the maximum \AbsEntropy\ of the two objects. Hence, it is easier to guess the absolute position of an object directly.

Not all correlation scenarios are easily exploitable. In fact, if the \CorrEntropy\ is sufficiently high, we can consider the object to be secure, even if it exhibits a low correlation with other objects. The problem arises when the correlation is so high that an address leak could enable bruteforcing the position of other objects within a reasonable timeframe, which, in our analysis, was considered to have 20 bits of entropy, as we explain in Section~\ref{sec:results}. When this occurs, we can identify a \textit{Correlation Path} that could potentially be exploited.
The most severe case of correlation presents 0 \CorrEntropy\ because the offset is fixed. This issue was exploited in the famous \texttt{off2libc} attack~\cite{marco2014effectiveness} and led to the introduction of the Effective Entropy concept into \ac{ASLR}-related discussion~\cite{DBLP:conf/uss/HerlandsHD14}.

\subsection{ASLR Hardware Limitations}
\label{subsec:hardware_limitations}

The underlying hardware plays an important role in the randomization of memory objects. Theoretically, the maximum entropy for an object is determined by the address size, which is 64 bits or 32 bits according to the hardware architecture. However, the practical limit is lower because of \textit{Multilevel Paging} and \textit{memory pages}.

Multilevel Paging is a technique that translates virtual memory addresses into physical ones. The idea is to split the address into several parts, each serving as an index into a different level of the page table hierarchy. For instance, in Intel's 4-level paging, an address is divided into four segments of 9 bits each, plus a 12-bit page offset (12 \textit{Least Significant Bits} (\texttt{LSb})), for a total of 48 bits. Consequently, the 16 \textit{Most Significant Bits} (\texttt{MSb}) are not supported at the hardware level -- they are represented as the sign extension of the address -- which leaves only 48 bits available for addressing memory. Moreover, even with the latest 5-level paging implementation, the practical limit for \ac{ASLR} remains at 56 bits.

On the other hand, memory pages are the smallest unit of memory that a kernel can allocate, and they are always allocated aligned with the page's size. Therefore, the \texttt{LSb} of an address represent the offset within the page, and they cannot be used in the randomization process. For instance, in Intel x86\_64 systems we have \texttt{4KB} pages, and thus the 12 \texttt{LSb} are used for the page offset, while in ARM M1 systems (64 bits) we have \texttt{16KB} pages~\cite{DBLP:conf/uss/YuDJKF23}, and thus the 14 \texttt{LSb} are used for the page offset. Hence, the bits that can be used for randomization are $48 bits - 12 bits = 36 bits$ for Intel x86\_64 systems and $48 bits - 14 bits = 34 bits$ for ARM M1 systems.

In 32-bit systems, Multilevel Paging covers all of the bits of an address, but we still have \texttt{4KB} pages. Therefore, the maximum entropy for an object is $32 bits - 12 bits = 20 bits$, which is relatively low, making \ac{ASLR} less effective in 32-bit systems. For this reason, we exclude 32-bit systems from our analysis.

\subsection{ASLR Base Addresses}
\label{subsec:aslr_base_address}

Other factors that can weaken \ac{ASLR} effectiveness are \textit{\ac{ASLR} base addresses}. An \ac{ASLR} base address is an address, chosen either at runtime or boot-time, used to allocate a sequence of memory objects, either after or before the base address. These objects are allocated by applying an offset to the previously allocated object, and the offset can be one of the following: \textit{zero}, \textit{fixed}, \textit{alignment}, or \textit{random}. 

When the offset is zero, objects are allocated contiguously. This is common in most \acp{OS} when allocating executables, where all sections (e.g., \texttt{.text} and \texttt{.bss}) are placed contiguously to maintain cross-references and improve performance. A fixed offset, typically of at least one page, is often used for guard pages, like those in the Scudo~\cite{android_scudo} Android allocator, to prevent buffer overflows. An alignment offset is used to align pages of different sizes, such as aligning a huge page to its size. Finally, a random offset, typically multiple of a page's size, is used to allocate an object at a random position from the previous one and is the only offset that can increase the \CorrEntropy\ between objects. However, if the random offset is too small, the \CorrEntropy\ remains low, allowing attackers to reduce the search space for correlated objects.

For instance, in the latest Linux distributions, we have three main \ac{ASLR} base addresses: the executable base address, the shared libraries base address, and the stack base address. The executable base address is used for both the executable and the heap. After the executable is allocated, the kernel applies a random offset to its end address to compute the heap address, which is retrieved by the application via the \texttt{brk(NULL)} system call. The shared libraries base address handles shared libraries and dynamic memory pages allocated through \texttt{malloc()} (when the size exceeds \texttt{M\_MMAP\_THRESHOLD}) or \texttt{mmap()}. Here, memory pages are usually allocated contiguously, or with small random offsets, towards lower addresses, resulting in low \CorrEntropy\ that can potentially lead to Ret2Libc or Ret2LD attacks~\cite{marco2014effectiveness}. Finally, the stack base address is used to allocate the stack, arguments, and environment variables. At program startup, the kernel allocates environment variables and arguments at the stack's top, and afterward, the userspace application allocates functions' frames. Similar to the shared libraries base address, allocations are contiguous with a random offset between environment variables and arguments, once again resulting in low \CorrEntropy.

Although identifying \ac{ASLR} base addresses may be challenging due to missing documentation or code availability (e.g., in Windows), similar situations occur in other \acp{OS}. Generally, the stack base address is present across all \acp{OS}, while others could be different as they may allocate different objects. Despite potentially weakening \ac{ASLR} effectiveness, base addresses are a common practice across \acp{OS} to improve performance and simplify \ac{ASLR} implementations. In Section~\ref{sec:results}, we will see that most correlated objects are due to the \ac{ASLR} base addresses.

\subsection{OSes Choice Motivation}
\label{subsec:os_choice}

Looking at research related to \ac{ASLR} analysis, we can see a stable trend toward Linux systems. This is justified by the predominance of Unix servers active in 2023, counting for around 80\% of the market share, of which around 50\% uses Linux kernels~\cite{W3Techs_OS_Server}. Enterprise servers require higher security standards than consumer systems, so, understandably, those have received more attention, sometimes evaluating even hardened versions of the Linux kernel available on the market. On the other hand, we have the consumer market, where Linux-based systems count for 3\% of the market share when compared with 70\% of Windows and 20\% of MacOS systems~\cite{StatCounter_Desktop}. Moreover, the recent adoption of ARM architectures by Apple made all the research regarding MacOS outdated since they focus only on  x86\_64 architectures and, as discussed in Section~\ref{subsec:hardware_limitations}, the virtual address translation is handled differently~\cite{apple_arm_translation_code,DBLP:conf/uss/YuDJKF23}. Therefore, we decided to focus our research on the following operating systems: Windows (version 11), macOS (Ventura 13.4.1), and Linux (Ubuntu 22.04). We excluded Android from our research, despite its Linux kernel base, as we focused solely on desktop systems. Moreover, Android presents inherent challenges that require more in-depth analysis, such as variability in memory page sizes, differences across devices due to vendor-specific customizations, the non-deterministic Scudo heap allocator, and the difficulty in accessing memory objects managed by the Android Runtime (ART) environment. For a more detailed explanation, we refer the reader to \autoref{sec:android}

Finally, as discussed in Section~\ref{subsec:hardware_limitations}, hardware architecture plays a crucial role in the randomization of memory objects. Thus, we consider the x86\_64 architecture for Windows and Linux, and the ARM M1 architecture for macOS. For Linux, we consider two kernel versions (5.17.15 and 6.4.9) to study the impact of Memory Folios on Linux ASLR implementation. For macOS, we consider both the native ARM M1 architecture and the Rosetta dynamic binary translator, which allows applications compiled for x86\_64 to run on Apple ARM processors like the M1 chip.

\subsection{Threat Model}
\label{subsec:threat_model}

We consider two exploitation scenarios: \ding{192} \textit{Local Exploitation} and \ding{193} \textit{Remote Exploitation}. In the first scenario, the attacker has access to the target system, either remotely or physically, and can run arbitrary code. The attacker thus knows the \ac{OS}, and the goal is to leak sensitive data or to perform a \textit{privilege escalation}, i.e., to obtain superuser privileges. In the second scenario, the attacker does not have access to the target system but can interact with it through an exposed service. In this scenario, the attacker aims to achieve one or more of the following goals: \textit{Remote Code Execution (RCE)}, \textit{Privilege Escalation}, or \textit{Data Leakage}. In addition, the attacker does not know the \ac{OS} nor its version and must gather information about the system to exploit it or try exploits for different \acp{OS} and versions. 

In both scenarios, the attacker aims to exploit a memory corruption vulnerability, such as a buffer overflow or a use-after-free, on a userspace application to achieve the aforementioned goals. Hence, depending on the application and the vulnerability, the attacker may corrupt data structures, function pointers, and flow-related variables in the virtual memory of the target process. In general, the attacker needs to know the position of relevant memory objects in order to perform the exploitation. For instance, whenever the attacker can change the return address of a function, performing a \textit{Return-Oriented Programming (ROP)} attack, the attacker needs to know the position of the gadgets in the memory. These gadgets can be found in the executable and in shared libraries. Therefore, the attacker either has a \textit{memory leak} vulnerability that reveals the positions of these objects or needs to guess their positions. As we will see in Section~\ref{sec:results} and Section~\ref{sec:attacks_to_aslr}, the attacker may reduce the guessing space with a memory leak of another correlated object. It is important to note that the relevance of a memory object depends on the specific vulnerability and the target application. Most of the time, the attacker is interested in the executable and the libraries, as they contain code to perform control flow hijacking. However, in some cases, the attacker may be interested in other objects, like the heap, to perform a \textit{Heap Spraying} attack, or the stack, to target flow-related variables and hijack the flow of the application.

When the exploit fails because the object is not in the guessed position, we usually have a \textit{Segmentation Fault (SEGFAULT)} signal, and the process is terminated. In the local scenario, the userspace application can be restarted, and the attacker can try the exploit again. In the remote scenario, the exposed services usually have a parent process that waits for incoming connections and delegates the connection to a child process. Therefore, in case of a segmentation fault, only the child process is terminated, and the attacker can try the exploit again by opening another connection.

Finally, we assume an attacker capable of performing 300 \textit{tries per second (tps)}, i.e., the number of exploit attempts an attacker can perform in a second~\cite{marco2019address,Gisbert2016ExploitingLA_PositiveEnt}. We consider 300 tps as a reference value as, at this rate, an attacker can guess the position of a memory object with an entropy of \textbf{20 bits} in approximately 1 hour. This value can be more or less realistic depending on the target application logic, the hardware on which the application runs, and the exploit's size. On consumer-grade hardware, we observed tps ranging from 30 to 500 with one core. In Section~\ref{sec:attacks_to_aslr}, we will provide an instance of a real-world application where we achieve approximately 300 tps with an exploit running on a single core and more than 1,000 tps with the same exploit running on multiple cores. Remotely, estimating tps is more challenging due to additional factors like the proximity to the victim machine and the network speed~\cite{marco2014effectiveness}. However, 300 tps remains a realistic estimate even in remote scenarios, as attackers can parallelize the exploit across multiple threads, cores, and machines. Additionally, in a remote scenario, the attacker can perform the exploitation using machine(s) geographically close to the target one, reducing the network latency. As references, with 300 tps, an attacker can guess a memory object with 30 bits of entropy in around 41 days, a memory object with 25 bits of entropy in around 1 day, a memory object with 20 bits of entropy in around 1 hour, and a memory object with 15 bits of entropy in around 2 minutes. 
\section{Related Works}
\label{sec:related_works}

The most advanced tool used in research is \textbf{ASLR-A} by \textit{Marco-Gisbert and Ripoll} \cite{marco2019address, Gisbert2016ExploitingLA_PositiveEnt}. It was used to perform analysis on Linux 4.15, PaX (a hardened version of Linux kernel), and MacOS (originally referred to as OS X). As mentioned before, we will consider only the consumer \acp{OS}, so PaX implementation is out of the scope of this research. 
The tool was developed to overcome the limitation found in \textit{paxtest}, a tool developed by the PaX team to evaluate the performance of their newly developed ASLR implementation. \textit{paxtest} had several issues. It considered only \AbsEntropy, using a custom heuristic not always accurate when dealing with non-uniform distributions. Moreover, the limited number of tests does not provide statistical significance to the results. Finally, the analysis was incorrect in some cases; the sampling of text area was, in reality, the library section.

They improved those aspects by developing ASLR-A, a tool capable of taking thousands of samples at a second and able to analyze numerous statistics. However, in this document, we focused mainly on two aspects that are the ones easily exploitable in bruteforce attacks: \textbf{\AbsEntropy} and \textbf{\CorrEntropy}. The tool can provide \AbsEntropy\ estimation using three different methods:
\textit{Shannon}, \textit{Shannon} at byte level, \textit{Shannon} with variable bins width, and \textit{bit-flipping}. Based on \cite{marco2019address}, it seems to be capable of estimating also \CorrEntropy.  However, the last known version of the tools available on the researcher's website~\cite{ASLRA_tool} provided only a correlation matrix without the estimation of \CorrEntropy. In the end, the tool provides a good insight into the Probability Distribution of sections. 

The only true limitation we can identify in this research is the limited scope of objects and \acp{OS} considered. In fact, as mentioned in the previous section, \ac{ASLR} performance is related to many runtime conditions, such as memory fragmentation, thread execution, and allocation patterns, so the allocation of multiple objects per section and multiple threads can lead to a change in randomization performance. The same considerations are valid for their MacOS analysis. However, it is not clear how the samples for this system were collected as the randomization is performed only at boot-time.
No other research is available on the MacOS platform.

To the best of our knowledge, there are only three published studies on Windows. The first, regarding Windows Vista \cite{whitehouse2007analysis}, is outdated, so we will focus only on the ones analyzing Windows 10 \cite{diaz2021address}  and Windows 7 \cite{aristizabal2013measuring}. The sampling of Windows 10 was performed through \texttt{5000} reboots using a custom-written tool, which took a total of \texttt{500,000} samples, while for the sections that were randomized at runtime \texttt{5 mln} samples were considered \cite{diaz2021address}. The results are not publicly available but, based on the researcher's claims, they were able to estimate the \AbsEntropy\ of memory objects, probability distribution, and their correlation; however, no mention of \CorrEntropy\ was made, and they just considered the main execution flow, without launching multiple threads; moreover, as in the case of ASLR-A, no attention to doing multiple allocations of different sizes where taken \cite{aslr_tool}.

The analysis of Windows 7 \cite{aristizabal2013measuring}, even if it is outdated and considered only four memory sections, concluded that the problems highlighted in Windows Vista \cite{whitehouse2007analysis} were still present: heap-allocated objects with non-uniform distribution and shared libraries randomized at boot-time.

Over the years, many pointed out that PRNG on Android has low entropy  \cite{DBLP:conf/icc/DingPZZ14, DBLP:conf/woot/KaplanKHD14}. Moreover, because every process is forked from Zygote, we can expect poor runtime performance of ASLR \cite{DBLP:conf/sp/LeeLWKL14}. Another problem of Android security is the customization made by vendors \cite{DBLP:conf/iscis/LiebergeldL13}. This aspect is hard to analyze due to the fragmentation of the Android hardware and vendors. Thus, we decided to focus the analysis on desktop \acp{OS} only.

\subsection{Limitations and Improvements}
\label{subsec:considerations_soa}

All mentioned researches have at least one of the following limitations: \ding{192} Lack of a broad \acp{OS} analysis, \ding{193} missing thread execution, \ding{194} inadequate sampling size, \ding{195} limited considered sections, \ding{196} limited allocations, \ding{197} missing \CorrEntropy\ estimation, or \ding{198} unclear entropy estimator choice.

This last point is strictly related to the inadequate sampling size. For example, to obtain an accurate estimation using direct Shannon entropy, we need $\mathbf{O\left(\frac{k}{\log(k)}\right)}$ samples where \textbf{k} is the number of symbols considered \cite{DBLP:journals/tit/AcharyaOST17}. As mentioned before, the maximum entropy obtainable on the considered system is 35 bits, therefore a k size of $\mathbf{2^{35}}$. To Estimate the entropy using the Shannon formula, we will need  \(\mathbf{\frac{2^{35}}{\log_{10}(2^{35})} = \mathbf{3.261.159.434}}\) samples which are way too much to be collected in a reasonable time. Even if we consider the best in class, Linux, which in some sections comes close to 30 bits of entropy, we are still considering hundreds of millions of samples to be collected. This problem is even more relevant when we take into consideration reboot times, so we need a less greedy estimator.

The use of Shannon at the byte level or other sorts of plug-in methods is a good approach, significantly reducing the number of samples needed to obtain a good estimation. However, they tend to overestimate the value of entropy due to outliers or due to non-uniform distribution.
The bit-flipping and other bit mask estimators are indicators of the changing bits of the address and only give a rough upper bound to the entropy value.

\section{How to Evaluate ASLR Implementations}
\label{sec:arch_and_met}

Assessing the effectiveness of \ac{ASLR} implementations is a challenging task due to the dynamic nature of virtual memory allocations in a program. These allocations can occur at various stages, such as boot-time, program startup, or runtime. Therefore, to understand how these memory allocations are related, it is necessary to conduct an empirical analysis. This is done by collecting the memory addresses of different objects and sections of the program, over multiple runs, and then analyzing the randomness of the memory layout and the correlation between different objects. In this way, through the statistical  analysis of millions of samples, we evaluate the strengths and weaknesses of \ac{ASLR} implementations.

\subsection{Sampling}
\label{subsec:sampling}

For the sampling phase, the main focus is efficiency and granularity of information. We wanted to emulate the behavior of real-world software to provide information about the effort needed to hit a specific object in memory and not only the page it belongs to. On the one hand, we need as much data as possible to better analyze the \ac{ASLR} performance across the different operating systems. On the other hand, the resources at our disposal are limited. Fortunately, we can define a unique subset of objects and allocation sizes, shared across all considered platforms, able to provide a solid picture of the \ac{ASLR} details in an efficient and homogeneous way.

\mypar{Memory Objects}
\label{subsubsec:mem_objects}
Facing the problem of choosing which memory object and section to sample in our research, we decided to focus on the interactions and correlations between objects; as a consequence, our sampling program makes multiple allocations of different sizes from 3 different flows: two independent threads (ThA, ThB) and the main thread (M). Because of this, the total number of addresses collected with each sample is around 60 (accounting for some platform limitations). Additionally, for Executable and Linkable Format (ELF) and Portable Executable (PE), we have different sections such as \texttt{.text}, \texttt{.bss}, and \texttt{.data}. In these cases, we do not need to sample all the sections, as they are placed contiguously in the virtual memory of the process. As a result, we only need to sample one section to know exactly the position of all the other sections. Finally, for simplicity, we group all these sections under a single memory object called \texttt{executable}.

To represent our memory objects, we use the syntax \textit{<object>\-\_\-<thread>}, where \textit{object} is the name of the object, and \textit{thread} is the thread to which the object belongs. For instance, \textit{stack\_M} is the stack of the main thread. For dinamically allocated object instead, we use the syntax \textit{<function\_name>\_<size>\_<thread>\_<sequence\-\_\-number>}, where \textit{<function\_name>} is the function that allocates the object, \textit{<size>} is the requested size, and \textit{<sequence\_number>} is the i-th operation of that type. For instance, \textit{malloc\_4KB\_ThB\_2} is the second 4-kilobytes malloc function call performed by thread B.

\begin{table}
    \centering 
    \caption{Selected Sizes for \texttt{malloc()} allocations.}
    \label{table:sizes_reduced}
    \small
    \resizebox{\columnwidth}{!}{
        \begin{tabular}{c|c|c|c|c|c|c|}
            \hhline{~------} 
            \rowcolor{gray!30}
            \cellcolor{white}                                                   & \textbf{16B}                                                      & \textbf{512B}                         & \textbf{4KB}                          & \textbf{256KB}                                                            & \textbf{4MB}                              & \textbf{128MB}        \\
            \hline
            \multicolumn{1}{|c|}{\cellcolor{gray!10}\textbf{Linux 6.4}}         & \multicolumn{3}{c|}{\cellcolor{green!10}\texttt{[heap] brk()}}                                                                                    & \cellcolor{orange!20}\texttt{pages}                                       & \multicolumn{2}{c|}{\cellcolor{red!20}\texttt{folio pages}}       \\
            \hline
            \multicolumn{1}{|c|}{\cellcolor{gray!10}\textbf{Linux 5.17}}        & \multicolumn{3}{c|}{\cellcolor{green!10}\texttt{[heap] brk() }}                                                                                   & \multicolumn{3}{c|}{\cellcolor{orange!20}\texttt{pages}}                                                                                      \\
            \hline
            \multicolumn{1}{|c|}{\cellcolor{gray!10}\textbf{MacOS M1 13.4.1}}    & \cellcolor{orange!10}\texttt{M\_NANO}                             & \cellcolor{blue!10}\texttt{M\_TINY}   & \cellcolor{pink!50}\texttt{M\_SMALL}  & \multicolumn{2}{c|}{\cellcolor{yellow!30}\texttt{M\_MEDIUM}}              & \cellcolor{purple!30}\texttt{M\_LARGE}                            \\
            \hline
            \multicolumn{1}{|c|}{\cellcolor{gray!10}\textbf{Windows 11}}        & \multicolumn{4}{c|}{\cellcolor{lime!20}\texttt{[heap]}}                                                                                                                                                                       & \multicolumn{2}{c|}{\cellcolor{orange!20}\texttt{pages}}          \\
            \hline
        \end{tabular}
    }
\end{table}

\mypar{Allocation Size Choice}
\label{subsubsec:size_choice}
Operating systems often employ various allocation methods for different memory sizes to enhance performance. For instance, Linux uses a default threshold of \texttt{128KB}(\texttt{M\-\_\-MMAP\-\_\-THRESHOLD}) to decide whether to use the legacy system \texttt{brk()} or the \texttt{mmap()} function as an allocation method; moreover, this threshold is variable and is optimized at runtime based on the allocation pattern \cite{GlibcMalloc} of the program so we cannot take that for granted. Since the introduction of Folios in Linux 5.18 \cite{linux_folios}, we have one more variation of allocation method for \texttt{mmap()} of sizes \texttt{>2MB}, with a significant impact on randomization entropy, as we will see in Section~\ref{subsec:linux} As a consequence, we consider two possible allocation methods: the \texttt{malloc()} and the \texttt{mmap()} functions. We believe they represent the most common allocation methods for our operating systems. In particular, Windows has an equivalent \texttt{mmap()} function called \texttt{VirtualAlloc()}. 

For \texttt{malloc()} allocations, we consider different sample sizes ranging from \texttt{16B} to \texttt{128MB}, as reported in Table~\ref{table:sizes_reduced}. In particular, in Windows and Linux, allocations of several kilobytes through the \texttt{malloc()} internally result in \texttt{mmap()}/\texttt{VirtualAlloc()} function calls. As a result, the kernels label these memory areas as general pages instead of heap pages. However, we consider them as heap pages since they are the result of their standard memory allocators.

Instead, for \texttt{mmap()}/\texttt{VirtualAlloc()} allocations, we consider specific sizes, which are characteristic of the \acp{OS}. Specifically, the term Single page refers to \texttt{4KB} pages for both Linux and Windows and for \texttt{16KB} pages for MacOS. This difference is due to the design implementations; Apple M1 chips are designed to allocate \texttt{16KB} to enhance runtime performance. Instead, the terms Huge pages and Large pages refer to \texttt{2MB} pages for Linux and Windows, respectively.

\subsection{Entropy Estimator}
\label{subsec:eentropy}

The most important quantitative parameter we want to analyze is Information Entropy, which is a concept introduced by Shannon \cite{DBLP:journals/bstj/Shannon48} in 1948 that measures the information contained in a data source. From another point of view, it is a way to measure the randomness. In the context of \ac{ASLR}, which is a system that incorporates the approach of "security through obscurity", entropy is directly linked to the effort needed to guess the current position of a memory object in terms of trials. Because of that, having a reliable way to estimate the entropy is a crucial part of \ac{ASLR} analysis.

As mentioned in Section~\ref{subsubsec:weakness_low_entropy}, we are dealing with addresses the size of 47bit (127TB of addressed space), so exhaustively sampling all values of our source is practically impossible. Moreover, to use the Shannon Entropy estimator, we need more than one sample for each bin, so the number is even bigger. We are dealing with an under-sampled discrete source analysis, so we must use an estimator suited for this task. The most common method to estimate Shannon entropy is to consider each byte (or a subset of bytes of the address) as an \textit{independent random variable} and then combine the resulting entropy to estimate the one of the complete addresses (a.k.a. \textit{Plugin Entropy}). Even if, in theory, it is a good method, we considered the assumption about the independence of bytes with regard to each other too strong to be stated generally true. To completely avoid this assumption, we decided to use a not-binned estimator.

The best option we identified is the \textit{Nemenman, Shafee, Bialek (NSB)} estimator \cite{DBLP:conf/nips/NemenmanSB01, nsb_website}, which is a coincidence-based estimator that also provides us with \textit{posterior standard deviation} to quantify the uncertainty in the estimation result. Thanks to this, it is still one of the best estimators for under-sampled sources, outperforming both \textit{Shannon Entropy} and \textit{Plugin Entropy} \cite{DBLP:journals/entropy/HernandezS19}. It has a bias of $\frac{2^{S/2}}{N}$ \cite{nsb_website} where $S$ is the unknown entropy and $N$ is the number of samples. Thus, we can calculate the number of samples we will need in the worst-case scenario to have a bias of less than 5\%.

This method is also suitable for the \CorrEntropy\ estimation. For correlated objects, we observe that the resultant entropy typically falls below the minimum entropy observed in either object, thereby incurring a bias comparable to or less than that encountered in absolute entropy measurements. Conversely, when evaluating two independent objects, the \CorrEntropy\ may exceed the individual maximum entropies, necessitating a larger sample size due to the increased bias. Nonetheless, this bias is acceptable since our primary interest is to correctly quantify high Correlation Entropies. Consequently, precise \CorrEntropy\ values for independent objects are not a priority.

\subsection{Sample Collection}
\label{subsec:sample_dec}

Following the rule presented in Section~\ref{subsec:eentropy}, we consider the maximum theoretical entropy (i.e., 35 bits) to identify the minimum number of samples. In this case, we want to collect enough samples to have a bias lower than 5\%. Hence, $\frac{2^{35/2}}{N} < 0.05$. Solving for $N$, we find that we need at least 3,800,000 samples.

To collect the required samples, we employ one x86\_64 machine and one ARM M1 machine. Additionally, to collect Windows 11 samples, we employ a \ac{VM} running on Ubuntu 22.04. Such a \ac{VM} is virtualized to directly access the hardware and avoid any possible bias of the guest \ac{OS}. However, we have memory objects that are randomized at boot-time, and thus, we need to reboot the system to collect the samples of these memory objects. Due to the significant amount of reboots required and the slow boot-time of the \ac{VM}, it is not possible to collect the required amount of samples in a reasonable time. Hence, we first identify the \textit{Changing Bitmasks} -- i.e., how many bits change in the address of objects in memory over different runs or reboots -- to estimate the minimum number of samples to have a bias lower than 5\%. It is important to recall that the NSB estimator provides the posterior standard deviation to quantify the uncertainty in the estimation result. Hence, after performing the analysis on the required samples, we evaluate the uncertainty of the estimation to confirm that the bias remains below 5\%. In other words, we verify that the estimated \textit{Changing Bitmasks} is therefore correct.

The thresholds for runtime and boot-time randomized objects are reported in Table~\ref{table:os_runtime_threshold} and Table~\ref{table:os_boot-time_threshold} respectively, where T stands for Theoretical and CB for Changing Bitmask.

\begin{table}
    \caption{Thresholds for runtime randomized sections.}
    \label{table:os_runtime_threshold}
    \centering
    \small
    \begin{tabular}{|c|c|c|}
        \hline
        \textbf{\ac{OS}} & \textbf{Entropy (bits)} & \textbf{Min Samples}\\
        \hline
        Ubuntu & 35 (T) & 3,800,000 \\
        \hline
        Windows & 35 (T) & 3,800,000 \\
        \hline
        MacOS & 19 (CB) & 15,000 \\
        \hline
    \end{tabular}

\end{table}
\begin{table}
    \centering
    \caption{Thresholds for boot-time randomized sections.}
    \label{table:os_boot-time_threshold}
    \small
    \begin{tabular}{|c|c|c|}
        \hline
        \textbf{\ac{OS}} & \textbf{Entropy (bits)} & \textbf{Min Samples} \\
        \hline
        Windows & 19 (CB) & 15,000 \\ 
        \hline
        MacOS (M1 Native)  & 16 (CB) & 5,000 \\ 
        \hline
        MacOS (M1 Rosetta)  & 15 (CB) & 3,600 \\ 
        \hline
    \end{tabular}

\end{table}

\section{Randomness Analysis}
\label{sec:results}

In this section, we present the results of the analysis as well as the findings on the different \acp{OS} with the different configurations. As discussed in Section~\ref{subsec:threat_model}, we consider \textbf{20 bits} of entropy as the reference threshold. This threshold is artificial as, in reality, there is no real value of "safeness", and it strongly depends on how many tps an attacker can perform. We consider good results to be everything above this value and bad results to be everything under. Moreover, we briefly discuss the strengths and weaknesses of each system considering: \ding{192} \textit{Allocation Layout}, \ding{193} \textit{Probability Distribution},  \ding{194} \textit{Absolute Entropy}, and \ding{195} \textit{Correlation Entropy}.

\begin{figure*}
    \centering
    \includegraphics[width=0.9\textwidth]{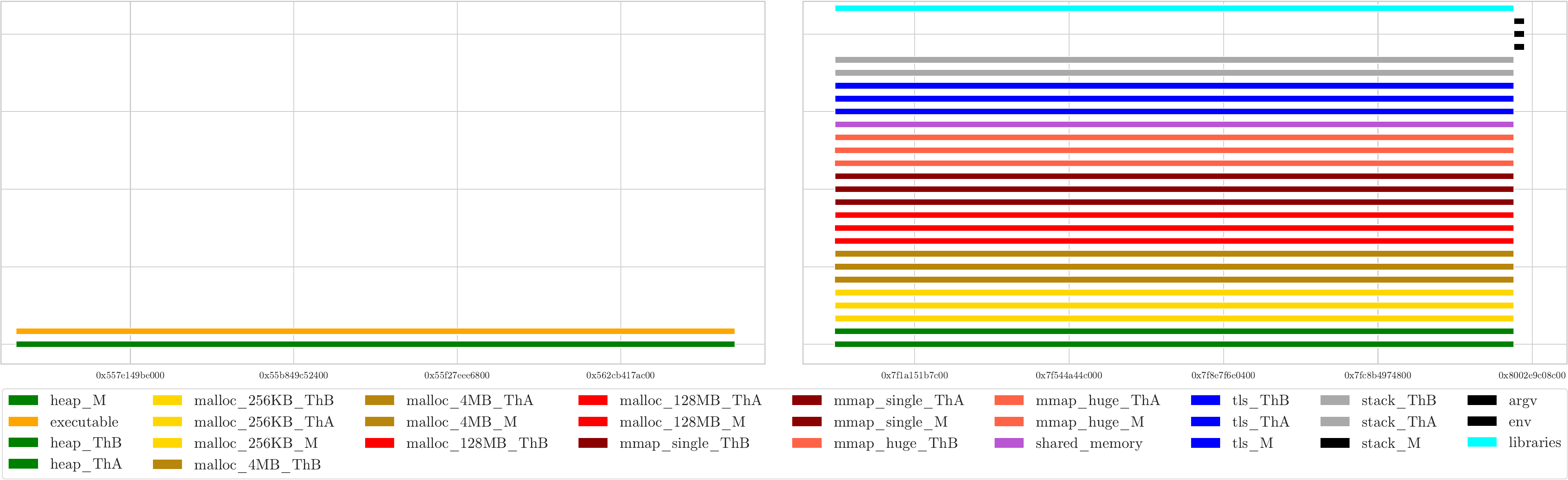}
    \caption{Linux (Ubuntu 22.04) Allocation Layout.}
    \label{fig:linux_layout}
\end{figure*}

To present \ding{192} \textit{Allocations Layouts}, we represent the different groups, sections, or objects as horizontal colored bars to qualitatively highlight in which range of addresses each object can be allocated. It is important to note that these layouts just represent possible allocation addresses and not how much they expand during the execution. To present \ding{193} \textit{Probability Distributions}, we rely on the \textit{Binned Histogram}, a discrete visualization method that represents the random nature of our memory objects. To present \ding{194} \textit{Absolute Entropies} results, we group objects and sections that are contiguous or with zero entropy to one another. These groups cannot be defined overall as they depend on the \ac{OS}. For instance, the glibc heap manager allocates all the chunks smaller than a few kilobytes on the same heap page. As a result, we consider all the small allocations to be a single group under Linux. Finally, to present \ding{195} \textit{Correlation Entropies} results, we report the \textit{Distance Entropy Matrices}. These matrices highlight the entropy of the distance between two objects or groups. The reader can find complete \CorrEntropy\ matrixes in the Appenidix~\ref{sec:corr_matrix_entropy}.
The complete list of figures is also available at: \url{https://zenodo.org/records/12968870}

\subsection{Linux}
\label{subsec:linux}

We perform our analysis on Ubuntu 22.04 running on an x86\_64 machine, employing the latest glibc library in Ubuntu 22.04, namely glibc 2.35. With the introduction of Folios in Linux kernel 5.18, we analyze two kernel versions: 5.17.15 and 6.4.9. We believe Linux Folios, as well as Huge pages (\texttt{2MB}), can drastically decrease the entropy of memory pages. As a result, we collected more than 4 million samples for each kernel version and evaluated their randomness.

\begin{table*}
    \centering
    \caption{Linux (Ubuntu 22.04) Absolute Entropy. \textit{lib\_small} is a library whose size is lower than 2MB. \textit{lib\_big} is a library whose size is bigger than 2MB. In our case, \textit{lib\_big} is the \textit{glibc}. This distinction is important since Linux Folios are used only for the \textit{lib\_big} allocation. In fact, \textit{lib\_big} for \texttt{Linux 6.4.9} has a lower entropy than \textit{lib\_small.}}
    \small
    \resizebox{\textwidth}{!}{
        \begin{tabular}{lrr||lrr||lrr||lrr}
            \multirow{2}{*}{\textbf{Object}}    & \multicolumn{2}{c||}{\textbf{Entropy}}                                  & \multirow{2}{*}{\textbf{Object}}    & \multicolumn{2}{c||}{\textbf{Entropy}}                                  & \multirow{2}{*}{\textbf{Object}}    & \multicolumn{2}{c||}{\textbf{Entropy}}                                  & \multirow{2}{*}{\textbf{Object}}    & \multicolumn{2}{c}{\textbf{Entropy}}                                    \\
                                                & \textbf{5.17.15}                   & \textbf{6.4.9}                     &                                     & \textbf{5.17.15}                   & \textbf{6.4.9}                     &                                     & \textbf{5.17.15}                   & \textbf{6.4.9}                     &                                     & \textbf{5.17.15}                   & \textbf{6.4.9}                     \\
            env                                 & \textcolor{black}{22.472}          & \textcolor{black}{22.472}          & heap\_M                             & \textcolor{green!60!black}{28.846} & \textcolor{green!60!black}{28.850} & heap\_ThA                           & \textcolor{red}{14.926}            & \textcolor{red}{14.978}            & heap\_ThB                           & \textcolor{red}{14.744}            & \textcolor{red}{14.931}            \\
            stack\&argv\_M                      & \textcolor{green!60!black}{30.832} & \textcolor{green!60!black}{30.836} & malloc\_256KB\_M\_\_1               & \textcolor{green!60!black}{28.845} & \textcolor{black}{25.832}          & malloc\_256KB\_ThA\_\_1             & \textcolor{black}{24.247}          & \textcolor{black}{23.998}          & malloc\_256KB\_ThB\_\_1             & \textcolor{black}{24.773}          & \textcolor{black}{24.006}          \\
            shared\_memory                      & \textcolor{green!60!black}{28.845} & \textcolor{green!60!black}{28.851} & malloc\_4MB\_M\_\_1                 & \textcolor{green!60!black}{28.845} & \textcolor{black}{19.611}          & malloc\_4MB\_ThA\_\_1               & \textcolor{black}{19.752}          & \textcolor{black}{20.007}          & malloc\_4MB\_ThB\_\_1               & \textcolor{black}{19.912}          & \textcolor{black}{20.197}          \\
            tls\_M                              & \textcolor{green!60!black}{28.845} & \textcolor{green!60!black}{28.836} & malloc\_128MB\_M\_\_1               & \textcolor{green!60!black}{28.845} & \textcolor{black}{19.611}          & malloc\_128MB\_ThA\_\_1             & \textcolor{red}{16.236}            & \textcolor{red}{16.071}            & malloc\_128MB\_ThB\_\_1             & \textcolor{red}{15.660}            & \textcolor{red}{16.134}            \\
            lib\_big                            & \textcolor{green!60!black}{28.845} & \textcolor{red}{19.023}            & mmap\_single\_M\_\_1                & \textcolor{green!60!black}{28.845} & \textcolor{green!60!black}{28.851} & mmap\_single\_ThA\_\_1              & \textcolor{green!60!black}{28.838} & \textcolor{green!60!black}{28.838} & mmap\_single\_ThB\_\_1              & \textcolor{green!60!black}{28.836} & \textcolor{green!60!black}{28.838} \\
            lib\_small                          & \textcolor{green!60!black}{28.845} & \textcolor{green!60!black}{28.851} & mmap\_huge\_M\_\_1                  & \textcolor{red}{19.024}            & \textcolor{red}{19.023}            & mmap\_huge\_ThA\_\_1                & \textcolor{red}{18.814}            & \textcolor{red}{18.875}            & mmap\_huge\_ThB\_\_1                & \textcolor{red}{18.869}            & \textcolor{red}{18.883}            \\
            executable                          & \textcolor{green!60!black}{28.862} & \textcolor{green!60!black}{28.840} & malloc\_256KB\_M\_\_2               & \textcolor{black}{25.911}          & \textcolor{black}{24.902}          & malloc\_256KB\_ThA\_\_2             & \textcolor{black}{23.722}          & \textcolor{black}{23.501}          & malloc\_256KB\_ThB\_\_2             & \textcolor{black}{24.154}          & \textcolor{black}{23.522}          \\
            stack\_ThA                          & \textcolor{red}{19.026}            & \textcolor{red}{19.023}            & malloc\_4MB\_M\_\_2                 & \textcolor{black}{19.588}          & \textcolor{red}{19.023}            & malloc\_4MB\_ThA\_\_2               & \textcolor{red}{18.929}            & \textcolor{red}{19.117}            & malloc\_4MB\_ThB\_\_2               & \textcolor{red}{18.977}            & \textcolor{red}{19.125}            \\
            tls\_ThA                            & \textcolor{red}{19.026}            & \textcolor{red}{19.023}            & malloc\_128MB\_M\_\_2               & \textcolor{black}{19.588}          & \textcolor{red}{19.023}            & malloc\_128MB\_ThA\_\_2             & \textcolor{red}{16.840}            & \textcolor{red}{16.786}            & malloc\_128MB\_ThB\_\_2             & \textcolor{red}{16.227}            & \textcolor{red}{16.636}            \\
            stack\_ThB                          & \textcolor{red}{19.026}            & \textcolor{red}{19.029}            & mmap\_single\_M\_\_2                & \textcolor{green!60!black}{28.845} & \textcolor{green!60!black}{28.851} & mmap\_single\_ThA\_\_2              & \textcolor{green!60!black}{28.836} & \textcolor{green!60!black}{28.843} & mmap\_single\_ThB\_\_2              & \textcolor{green!60!black}{28.835} & \textcolor{green!60!black}{28.836} \\
            tls\_ThB                            & \textcolor{red}{19.026}            & \textcolor{red}{19.029}            & mmap\_huge\_M\_\_2                  & \textcolor{red}{19.024}            & \textcolor{red}{19.023}            & mmap\_huge\_ThA\_\_2                & \textcolor{red}{18.711}            & \textcolor{red}{18.804}            & mmap\_huge\_ThB\_\_2                & \textcolor{red}{18.787}            & \textcolor{red}{18.808}            \\
        \end{tabular}}
    \label{tab:abs_ent_linux}
\end{table*}

\begin{figure}
    \centering
    \includegraphics[width=0.9\columnwidth]{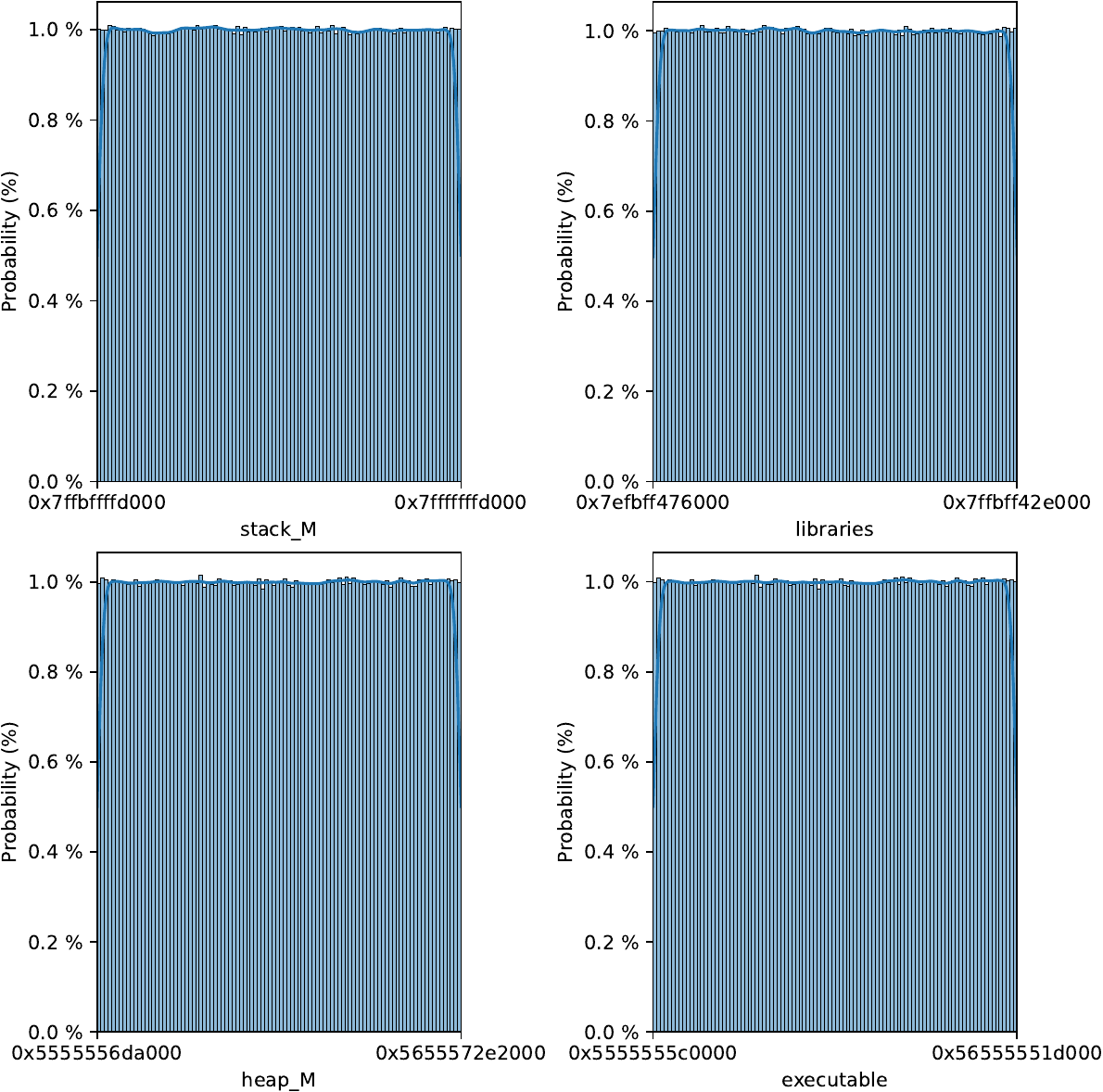}
    \caption{Probability Distribution in Linux (Ubuntu 22.04). All the distributions are available at: \url{https://zenodo.org/records/12968870/files/linux_distribution.pdf}}
    \label{fig:distrib_linux}
\end{figure}

\begin{figure*}
    \centering
    \includegraphics[width=0.9\textwidth]{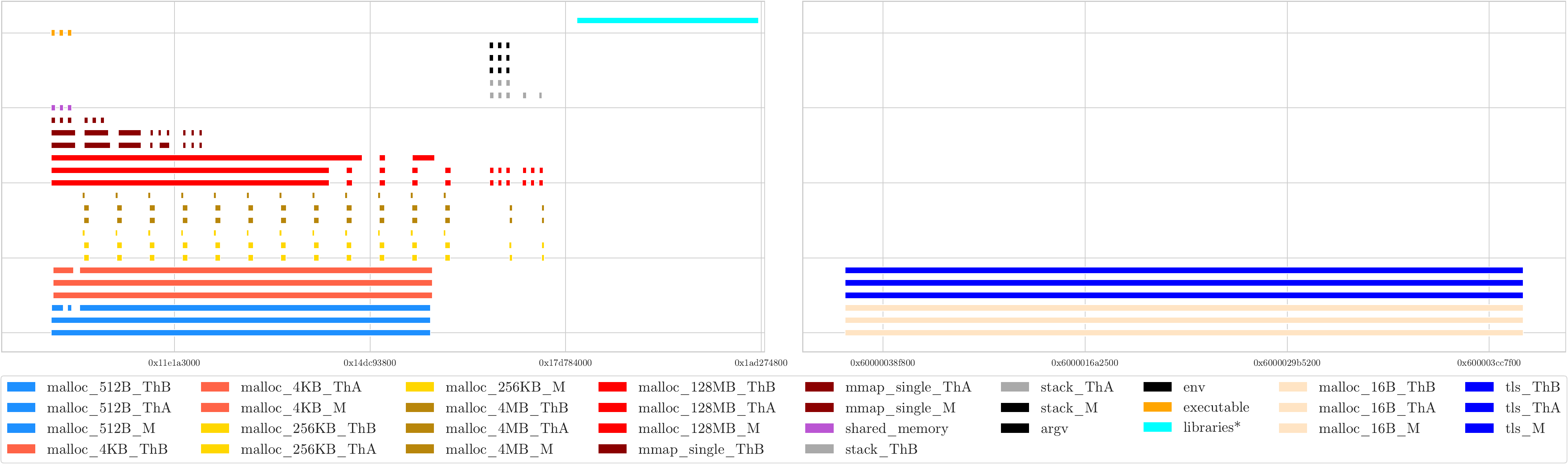}
    \caption{MacOS M1 Native Allocation Layout.}
    \label{fig:macosm1_layout}
\end{figure*}

\begin{figure*}
    \centering
    \includegraphics[width=0.9\textwidth]{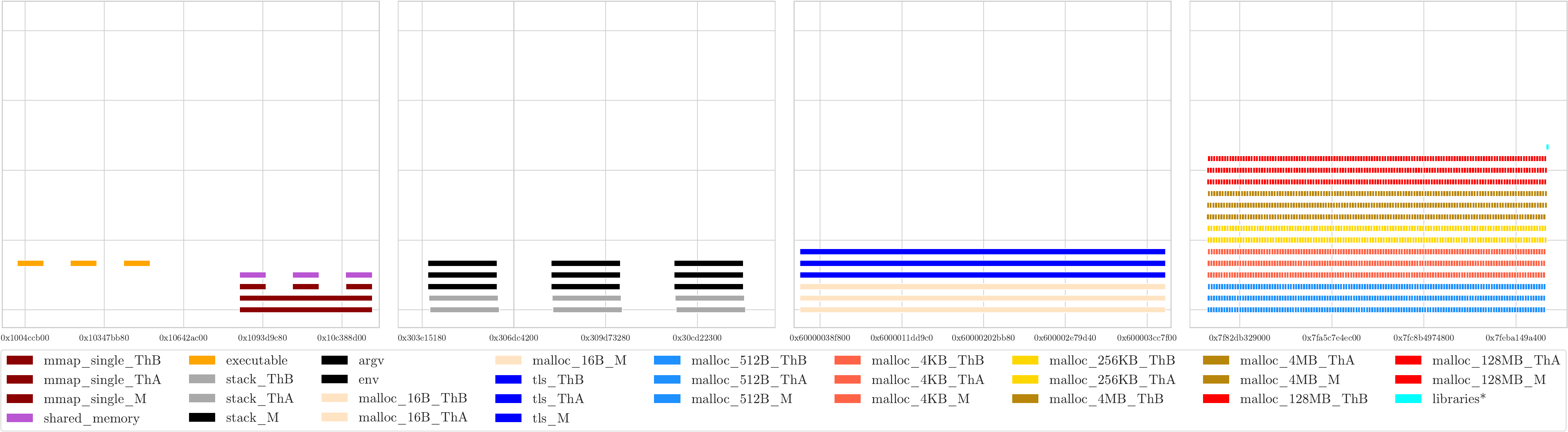}
    \caption{MacOS M1 Rosetta Allocation Layout.}
    \label{fig:macosm1_rosetta_layout}
\end{figure*}

All the samples have been collected without performing any reboot since the Linux kernel randomizes every memory object at runtime. Moreover, the sampling program is compiled with \textit{GCC} as \textit{Position Independent} (flag \texttt{-fPIE}) and relies on external libraries, i.e., not compiled with the flag \texttt{static}. Finally, the kernels are configured to fully randomize the objects in user space (\texttt{/proc/sys/kernel/randomize\_va\_space} = 2).

\mypar{Allocation Layout} As we can see in Figure~\ref{fig:linux_layout}, the allocation layout is the same for both kernel versions and is mainly divided into two regions with a large empty region in the middle. On the left, starting with lower addresses, we have the sections of the \texttt{executable} and the \texttt{heap}. This latter contains all the allocations of the main thread with sizes lower than a dynamic threshold. On the right of the figure, positioned among high memory addresses, we have the \texttt{stack}, the \texttt{libraries}, and all the allocations that are usually greater than \texttt{256KB}, allocated through the \texttt{mmap()} or the \texttt{malloc()} functions. Additionally, we can find other threads' \texttt{stack}, \texttt{heap} and \texttt{\ac{TLS}}. As you can notice, most of the object and group allocations almost completely overlap. This is due to the fact that some objects are allocated considering the allocation of other objects. For instance, the \texttt{heap} is allocated after the \texttt{exectuable}, considering a random offset starting from the end of the \texttt{exectuable}. This behavior is the leading cause of \CorrEntropy\ since the position of an object depends on another, especially when the random offset is limited. Similarly, the \texttt{libraries} are allocated considering a random offset starting from the end of the \texttt{stack}. In this case, the offset guarantees the \texttt{stack} to grow sufficiently, and this can be seen in the figure since the black bars are separated from the others.

\mypar{Probability Distribution} Regardless of the kernel version, Linux randomizes every object uniformly. This is evident from the probability distribution in Figure~\ref{fig:distrib_linux}. Figure~\ref{fig:distrib_linux} shows a histogram of 100 bins of the same size. The probability of a memory object being allocated in a specific range of addresses is the same for all the objects and is around 1\%, as expected for a uniform distribution. Moreover, the allocation range is also reported on the x-axis.

\mypar{Absolute Entropy} From Table~\ref{tab:abs_ent_linux} we can see the \AbsEntropy\ of both kernel versions. The most common and used sections of the main thread, such as the \texttt{stack}, the \texttt{executable}, the \texttt{heap}, and the \texttt{libraries}, achieve a relatively secure entropy of 28.8 bits, with the \texttt{stack} reaching up to 31.8 bits. However, in the kernel version 6.4.9, we have a drop in the entropy to 19 bits when the library is larger than a huge page (\texttt{2MB}). This is due to the Linux Folio optimization~\cite{folios_article,folios_git}. Linux Folio is a memory structure introduced to increase performance and reduce memory fragmentation, as it groups multiple consecutive memory pages of \texttt{4KB}, as a single bigger memory chunk. This uses neither huge pages nor transparent huge pages and is a flexible structure, so in theory, the size is not fixed. Its size is a power of two, and it is aligned with its size~\cite{linux_folios}, so for a Large Folio of \texttt{2MB}, we should see a page offset of 21 bits versus the 12 bits of a \texttt{4KB} page. As a consequence, all memory objects allocated using this new structure should expect around a 9-bit reduction in entropy. In fact, the sampled lib experiencing the reduction is  the \texttt{glibc} that is indeed bigger than \texttt{2MB}. This is a huge reduction for an executable memory section and gives an attacker almost four hundred times more chances of success compared to Linux version 5.17.15. Moreover, you can notice an entropy reduction after the \textit{mmap\_huge\_M\_\_1} in 5.17.15. This allocation indeed reduces the entropy of the next page allocations since they are relative to it; pages are allocated contiguously or almost contiguously and thus, a huge page determines the allocation of the next pages. In 6.4.9, a similar pattern is observed where the first huge page allocated is for \textit{lib\_big}, which in this case is \texttt{glibc}. We argue that having \texttt{glibc} easily guessable poses a serious security issue. Since \texttt{glibc} contains a lot of useful gadgets and functions, it is a common target for attackers to build exploits.

\begin{table*}
    \centering
    \caption{MacOS (Ventura 13.4.1) Absolute Entropy.}
    \small
    \resizebox{\textwidth}{!}{
        \begin{tabular}{lrr||lrr||lrr||lrr}
            \multirow{2}{*}{\textbf{Object}}    & \multicolumn{2}{c||}{\textbf{Entropy}}                                  & \multirow{2}{*}{\textbf{Object}}    & \multicolumn{2}{c||}{\textbf{Entropy}}                                  & \multirow{2}{*}{\textbf{Object}}    & \multicolumn{2}{c||}{\textbf{Entropy}}                                  & \multirow{2}{*}{\textbf{Object}}    & \multicolumn{2}{c}{\textbf{Entropy}}                                    \\
                                                & \textbf{Native}                    & \textbf{Rosetta}                   &                                     & \textbf{Native}                    & \textbf{Rosetta}                   &                                     & \textbf{Native}                    & \textbf{Rosetta}                   &                                     & \textbf{Native}                    & \textbf{Rosetta}                   \\
            env                                 & \textcolor{red}{12.903}            & \textcolor{red}{15.972}            & malloc\_16B\_M\_\_1                 & \textcolor{red}{12.581}            & \textcolor{red}{12.609}            & malloc\_16B\_ThA\_\_1               & \textcolor{red}{13.725}            & \textcolor{red}{13.372}            & malloc\_16B\_ThB\_\_1               & \textcolor{red}{14.050}            & \textcolor{red}{13.514}            \\
            stack\&argv\_M                      & \textcolor{red}{12.028}            & \textcolor{red}{15.121}            & malloc\_512B\_M\_\_1                & \textcolor{red}{7.551}             & \textcolor{red}{16.548}            & malloc\_512B\_ThA\_\_1              & \textcolor{red}{9.237}             & \textcolor{red}{17.710}            & malloc\_512B\_ThB\_\_1              & \textcolor{red}{9.394}             & \textcolor{red}{18.169}            \\
            shared\_memory                      & \textcolor{red}{11.583}            & \textcolor{red}{13.584}            & malloc\_4KB\_M\_\_1                 & \textcolor{red}{7.465}             & \textcolor{red}{16.502}            & malloc\_4KB\_ThA\_\_1               & \textcolor{red}{8.484}             & \textcolor{red}{17.412}            & malloc\_4KB\_ThB\_\_1               & \textcolor{red}{8.710}             & \textcolor{red}{17.628}            \\
            tls\_M                              & \textcolor{red}{12.549}            & \textcolor{red}{14.668}            & malloc\_256KB\_M\_\_1               & \textcolor{red}{3.189}             & \textcolor{red}{12.000}            & malloc\_256KB\_ThA\_\_1             & \textcolor{red}{4.550}             & \textcolor{red}{13.196}            & malloc\_256KB\_ThB\_\_1             & \textcolor{red}{4.755}             & \textcolor{red}{13.398}            \\
            libraries*                          & \textcolor{red}{15.625}            & \textcolor{red}{14.894}            & malloc\_4MB\_M\_\_1                 & \textcolor{red}{3.231}             & \textcolor{red}{12.059}            & malloc\_4MB\_ThA\_\_1               & \textcolor{red}{5.038}             & \textcolor{red}{13.600}            & malloc\_4MB\_ThB\_\_1               & \textcolor{red}{5.159}             & \textcolor{red}{13.942}            \\
            executable                          & \textcolor{red}{11.583}            & \textcolor{red}{13.583}            & malloc\_128MB\_M\_\_1               & \textcolor{red}{7.106}             & \textcolor{red}{16.070}            & malloc\_128MB\_ThA\_\_1             & \textcolor{red}{7.240}             & \textcolor{red}{16.093}            & malloc\_128MB\_ThB\_\_1             & \textcolor{red}{7.485}             & \textcolor{red}{16.095}            \\
            stack\_ThA                          & \textcolor{red}{11.583}            & \textcolor{red}{14.672}            & mmap\_single\_M\_\_1                & \textcolor{red}{11.587}            & \textcolor{red}{13.583}            & mmap\_single\_ThA\_\_1              & \textcolor{red}{12.478}            & \textcolor{red}{13.585}            & mmap\_single\_ThB\_\_1              & \textcolor{red}{12.454}            & \textcolor{red}{13.585}            \\
            tls\_ThA                            & \textcolor{red}{13.159}            & \textcolor{red}{14.333}            & malloc\_16B\_M\_\_2                 & \textcolor{red}{12.636}            & \textcolor{red}{12.692}            & malloc\_16B\_ThA\_\_2               & \textcolor{red}{14.516}            & \textcolor{red}{14.188}            & malloc\_16B\_ThB\_\_2               & \textcolor{red}{14.551}            & \textcolor{red}{14.573}            \\
            stack\_ThB                          & \textcolor{red}{11.583}            & \textcolor{red}{14.672}            & malloc\_512B\_M\_\_2                & \textcolor{red}{7.619}             & \textcolor{red}{16.615}            & malloc\_512B\_ThA\_\_2              & \textcolor{red}{9.984}             & \textcolor{red}{18.438}            & malloc\_512B\_ThB\_\_2              & \textcolor{red}{9.939}             & \textcolor{red}{19.102}            \\
            tls\_ThB                            & \textcolor{red}{13.519}            & \textcolor{red}{14.721}            & malloc\_4KB\_M\_\_2                 & \textcolor{red}{7.533}             & \textcolor{red}{16.579}            & malloc\_4KB\_ThA\_\_2               & \textcolor{red}{9.137}             & \textcolor{red}{18.032}            & malloc\_4KB\_ThB\_\_2               & \textcolor{red}{9.180}             & \textcolor{red}{18.489}            \\
                                                &                                    &                                    & malloc\_256KB\_M\_\_2               & \textcolor{red}{3.274}             & \textcolor{red}{12.128}            & malloc\_256KB\_ThA\_\_2             & \textcolor{red}{5.282}             & \textcolor{red}{13.844}            & malloc\_256KB\_ThB\_\_2             & \textcolor{red}{5.349}             & \textcolor{red}{14.256}            \\
                                                &                                    &                                    & malloc\_4MB\_M\_\_2                 & \textcolor{red}{3.278}             & \textcolor{red}{12.132}            & malloc\_4MB\_ThA\_\_2               & \textcolor{red}{5.295}             & \textcolor{red}{13.858}            & malloc\_4MB\_ThB\_\_2               & \textcolor{red}{5.360}             & \textcolor{red}{14.277}            \\
                                                &                                    &                                    & malloc\_128MB\_M\_\_2               & \textcolor{red}{7.857}             & \textcolor{red}{16.041}            & malloc\_128MB\_ThA\_\_2             & \textcolor{red}{6.650}             & \textcolor{red}{16.327}            & malloc\_128MB\_ThB\_\_2             & \textcolor{red}{7.162}             & \textcolor{red}{16.402}            \\
                                                &                                    &                                    & mmap\_single\_M\_\_2                & \textcolor{red}{12.046}            & \textcolor{red}{13.583}            & mmap\_single\_ThA\_\_2              & \textcolor{red}{12.611}            & \textcolor{red}{13.585}            & mmap\_single\_ThB\_\_2              & \textcolor{red}{12.582}            & \textcolor{red}{13.585}            \\
        \end{tabular}}
    \label{tab:abs_ent_macos}
\end{table*}

Further inspection of complete data in Table~\ref{tab:abs_ent_linux} highlights the low entropy of threads. We know from the documentation of \texttt{pthread\-\_\-create()}~\cite{pthread_create} that the default size of the thread stack is \texttt{2MB} ; being the thread stack allocated with \texttt{mmap()} function we see that the entropy of the \texttt{2MB} thread stack and the \texttt{4MB} \texttt{malloc()} are very similar, so this is the regular behavior. 

\begin{figure}
    \centering
    \includegraphics[width=0.9\columnwidth]{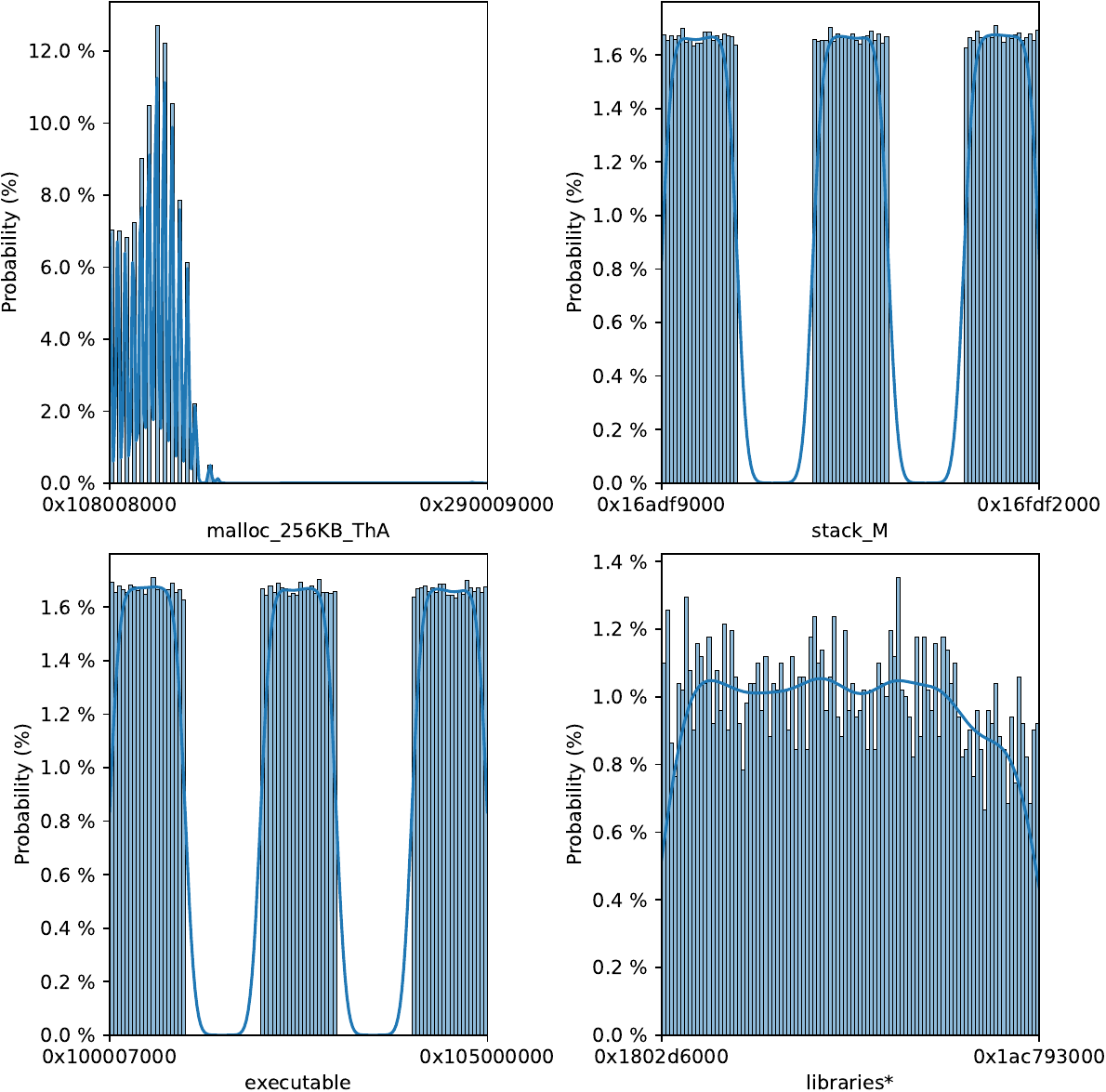}
    \caption{Probability Distribution in MacOS M1 Native. All the distributions are available at: \url{https://zenodo.org/records/12968870/files/macos_native_distribution.pdf}}
    \label{fig:distrib_macosm1_Native}
\end{figure}

\begin{figure}
    \centering
    \includegraphics[width=0.9\columnwidth]{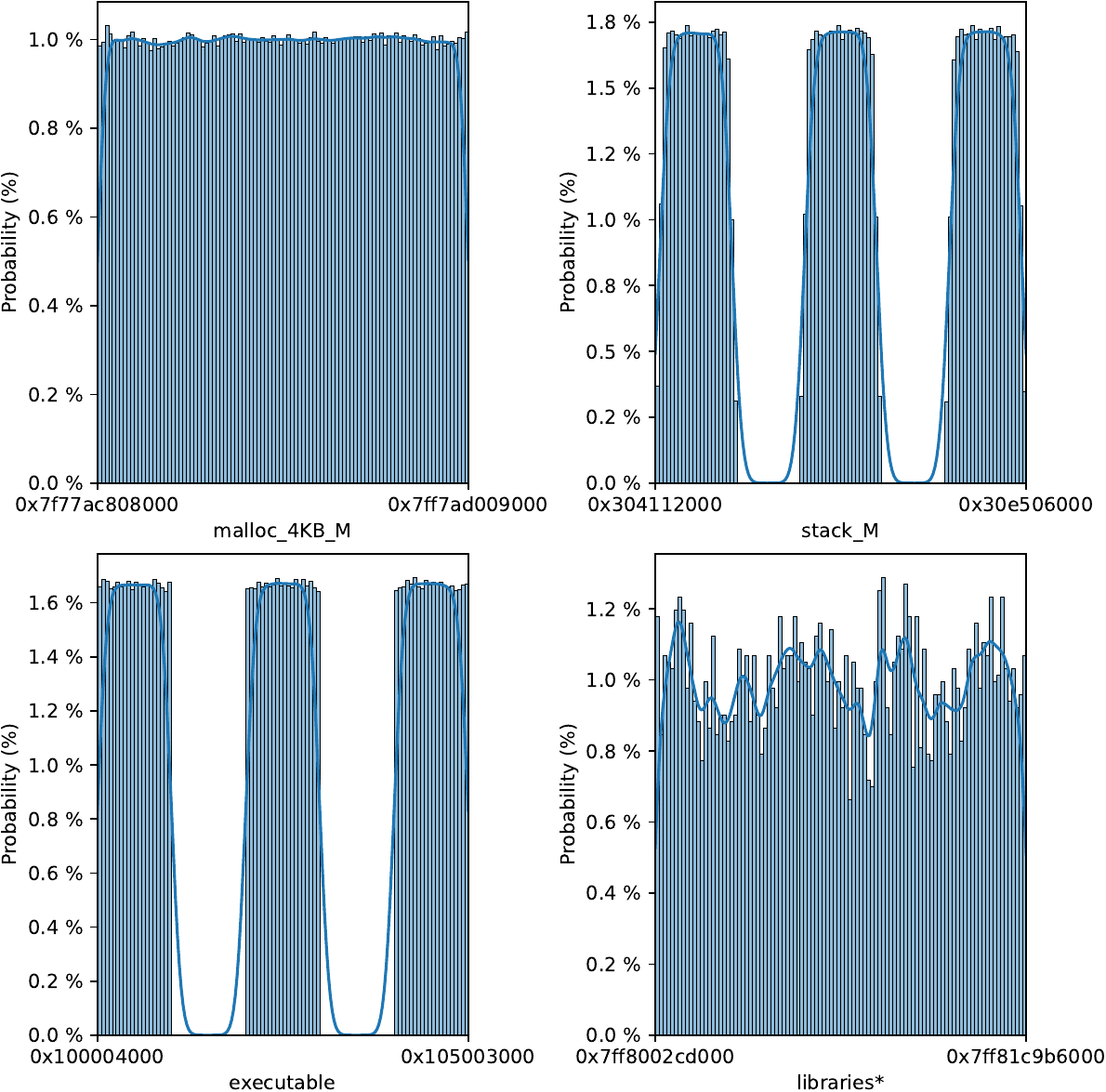}
    \caption{Probability Distribution in MacOS M1 Rosetta. All the distributions are available at: \url{https://zenodo.org/records/12968870/files/macos_rosetta_distribution.pdf}}
    \label{fig:distrib_macosm1}
\end{figure}

\mypar{Correlation Entropy} As expected from the allocation layout, many objects exhibit a high correlation with other objects. In particular, for both kernel versions, the \CorrEntropy\ between the executable and the heap is 13 bits, lowering their entropy by 15 bits; in other words, a leak can reduce the attack effort by 32,000 times. Moreover, all mapped areas suffer from consecutive allocation. Finally, libraries allocated with Linux Folios (\textit{lib\_big}) and libraries allocated with standard pages (\textit{lib\_small}) are no longer contiguous. Instead, they have 9 bits of \CorrEntropy. This is a positive side effect since the position of other libraries changes with every execution, but the \CorrEntropy\ is extremely low and can be easily bruteforced.

\subsection{MacOS}
\label{sec:results_macos_arm}

We analyze Mac OS M1 Ventura 13.4.1 on the Native Apple M1 (ARM) processor and on x86\_x64 using the Rosetta framework. MacOS is a system that randomizes libraries only at boot, so we reboot the machine several times to collect the samples. For each reboot, we collect 500 samples, and we restart the machine 5,500 times, thus obtaining 2,750,000 runtime samples and 5,000 boot-time samples. Additionally, the sampling program is compiled with \textit{Clang} as \textit{Position Independent} (flag \texttt{-fPIE}) and relies on external libraries, i.e., not compiled with the flag \texttt{-static}. Unlike Linux, macOS does not offer an option to adjust the level of randomization; therefore, we rely on the \ac{OS}'s default randomization settings.

\mypar{Allocation Layout} The allocation on MacOS using Native ARM architecture, visible in Figure~\ref{fig:macosm1_layout}, are mainly grouped among low addresses with only very small allocations that belong to the so-called \texttt{MALLOC\_NANO} area on the higher end of the memory. This suggests a high correlation between objects located at low addresses. In Figure~\ref{fig:macosm1_rosetta_layout} instead, we can see that the allocation layout on MacOS using Rosetta is divided into four regions. The lower one locates explicitly mapped pages, the executable, and the shared memories. The second one locates the stack of main and threads, while the third one corresponds to the higher of the Native system, accommodating \texttt{MALLOC\_NANO} and \texttt{TLS} variables. The higher one is dedicated to \texttt{malloc()} objects and libraries that, this time, present more visual distribution. Although we should expect a higher entropy, given the allocation ranges, the allocation of these objects is fragmented, which means that the alignment of these objects is much higher than a single page(\texttt{16KB}). This can result in a lower entropy.

\mypar{Probability Distribution} On MacOS using Native ARM architecture, there are some uniform allocations regarding the \texttt{TLS} objects and \texttt{MALLOC\_NANO} objects. Other sections have distributions characterized by high spikes or large groups, indicating a very low entropy. Figure~\ref{fig:distrib_macosm1_Native} shows the probability distribution of the memory objects. The histogram plot is divided into 100 bins; in the uniform distribution, the probability of each bin is 1\%. We can see in Figure~\ref{fig:distrib_macosm1_Native} spikes up to 14\% for malloc of \texttt{4MB}. Other objects (executable and stack) have a spotted uniform distribution. We can clearly see some gaps in the distribution; these are the probability of some addresses being higher than others.

In some sections, the randomization is probably linked to the intrinsic, not deterministic positioning of allocation and not to explicit \ac{ASLR} action.
Rosetta does a good job distributing the allocations, with all \texttt{malloc()} objects having a uniform distribution. The remaining objects have a uniform distribution that is not contiguous. The reader can see the mentioned in Figure~\ref{fig:distrib_macosm1}.
Finally, on both systems, libraries have a uniform distribution even if it is not visible due to the smaller amount of reboot samples.

\mypar{Absolute Entropy} As we can see from Table~\ref{tab:abs_ent_macos}, the overall entropy is low, with poor randomization on executable objects. Libraries have an insufficient boot-time randomization entropy (12.2 bit), while the executable has a worse runtime randomization entropy (11.5 bit) in Native ARM and slightly better entropy (13.5) using Rosetta. Moreover, with the Native system, the objects allocated with \texttt{malloc()} are very poorly randomized, ranging from a maximum entropy of 12 bits to a minimum entropy of 3 bits for the main thread. Other objects and \texttt{mmap()} allocated objects achieve a lower entropy around 12 bit as well, which is still relatively low. Finally, thread allocation entropies are slightly better than the main thread ones. Instead, Rosetta entropies are overall higher but still insufficient and again particularly low in objects allocated with \texttt{malloc()} and \texttt{mmap()}. One reason for such a low entropy is the size of Apple M1 single pages. In fact, \texttt{16KB} pages require an alignment of 14 bits, i.e., only 33 bits instead of 35 can be used to randomize an address. Moreover, the allocations slots for these allocations are surprisingly at \texttt{1} to \texttt{4MB} (20 to 22 bits) of distance from one another, which confirms our hypothesis in Figure~\ref{fig:macosm1_rosetta_layout}.

\begin{figure*}
    \centering
    \includegraphics[width=0.9\textwidth]{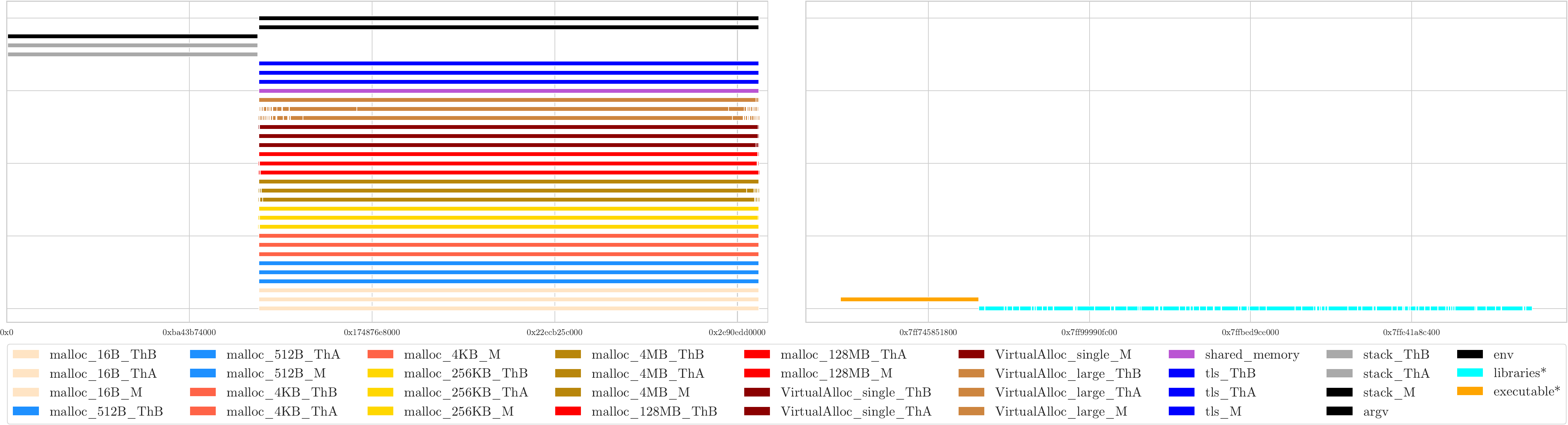}
    \caption{Windows 11 Allocation Layout.}
    \label{fig:win_layout}
\end{figure*}

\begin{table*}
    \centering
    \caption{Windows 11 Absolute Entropy.}
    \resizebox{0.9\textwidth}{!}{
        \begin{tabular}{lr||lr||lr||lr}
            \textbf{Object}                     & \textbf{Entropy}                      & \textbf{Object}         & \textbf{Entropy}                   & \textbf{Object}         & \textbf{Entropy}                   & \textbf{Object}           & \textbf{Entropy}                   \\
            env                                 & \textcolor{black}{25.914}             & malloc\_16B\_M\_\_1     & \textcolor{green!60!black}{30.818} & malloc\_16B\_ThA\_\_1   & \textcolor{green!60!black}{30.921} & malloc\_16B\_ThB\_\_1     & \textcolor{green!60!black}{30.822} \\
            argv                                & \textcolor{black}{25.416}             & malloc\_512B\_M\_\_1    & \textcolor{black}{25.914}          & malloc\_512B\_ThA\_\_1  & \textcolor{green!60!black}{31.558} & malloc\_512B\_ThB\_\_1    & \textcolor{green!60!black}{30.837} \\
            stack\_M                            & \textcolor{green!60!black}{28.151}    & malloc\_4KB\_M\_\_1     & \textcolor{black}{25.914}          & malloc\_4KB\_ThA\_\_1   & \textcolor{green!60!black}{28.961} & malloc\_4KB\_ThB\_\_1     & \textcolor{green!60!black}{30.368} \\
            shared\_memory                      & \textcolor{black}{25.345}             & malloc\_256KB\_M\_\_1   & \textcolor{black}{25.359}          & malloc\_256KB\_ThA\_\_1 & \textcolor{black}{25.881}          & malloc\_256KB\_ThB\_\_1   & \textcolor{black}{26.879}          \\
            tls\_M                              & \textcolor{black}{25.359}             & malloc\_4MB\_M\_\_1     & \textcolor{green!60!black}{29.250} & malloc\_4MB\_ThA\_\_1   & \textcolor{black}{24.755}          & malloc\_4MB\_ThB\_\_1     & \textcolor{black}{24.656}          \\
            libraries*                          & \textcolor{red}{18.966}               & malloc\_128MB\_M\_\_1   & \textcolor{black}{25.528}          & malloc\_128MB\_ThA\_\_1 & \textcolor{black}{23.057}          & malloc\_128MB\_ThB\_\_1   & \textcolor{black}{23.153}          \\
            executable*                         & \textcolor{red}{16.985}               & VirtualAlloc\_single\_M\_\_1    & \textcolor{black}{25.243}          & VirtualAlloc\_single\_ThA\_\_1  & \textcolor{black}{25.158}          & VirtualAlloc\_single\_ThB\_\_1    & \textcolor{black}{25.185}          \\
            stack\_ThA                          & \textcolor{green!60!black}{27.407}    & VirtualAlloc\_large\_M\_\_1      & \textcolor{black}{19.654}          & VirtualAlloc\_large\_ThA\_\_1    & \textcolor{red}{19.002}            & VirtualAlloc\_large\_ThB\_\_1      & \textcolor{red}{19.102}            \\
            tls\_ThA                            & \textcolor{green!60!black}{30.808}    & malloc\_16B\_M\_\_2     & \textcolor{green!60!black}{30.796} & malloc\_16B\_ThA\_\_2   & \textcolor{green!60!black}{31.852} & malloc\_16B\_ThB\_\_2     & \textcolor{green!60!black}{31.845} \\
            stack\_ThB                          & \textcolor{green!60!black}{27.407}    & malloc\_512B\_M\_\_2    & \textcolor{green!60!black}{28.872} & malloc\_512B\_ThA\_\_2  & \textcolor{green!60!black}{31.311} & malloc\_512B\_ThB\_\_2    & \textcolor{green!60!black}{31.487} \\
            tls\_ThB                            & \textcolor{green!60!black}{30.780}    & malloc\_4KB\_M\_\_2     & \textcolor{green!60!black}{28.931} & malloc\_4KB\_ThA\_\_2   & \textcolor{green!60!black}{28.937} & malloc\_4KB\_ThB\_\_2     & \textcolor{green!60!black}{30.751} \\
                                                &                                       & malloc\_256KB\_M\_\_2   & \textcolor{green!60!black}{28.931} & malloc\_256KB\_ThA\_\_2 & \textcolor{green!60!black}{27.407} & malloc\_256KB\_ThB\_\_2   & \textcolor{black}{26.749}          \\
                                                &                                       & malloc\_4MB\_M\_\_2     & \textcolor{black}{24.122}          & malloc\_4MB\_ThA\_\_2   & \textcolor{black}{23.838}          & malloc\_4MB\_ThB\_\_2     & \textcolor{black}{24.014}          \\
                                                &                                       & malloc\_128MB\_M\_\_2   & \textcolor{black}{23.340}          & malloc\_128MB\_ThA\_\_2 & \textcolor{black}{22.394}          & malloc\_128MB\_ThB\_\_2   & \textcolor{black}{22.538}          \\
                                                &                                       & VirtualAlloc\_single\_M\_\_2    & \textcolor{black}{25.209}          & VirtualAlloc\_single\_ThA\_\_2  & \textcolor{black}{25.079}          & VirtualAlloc\_single\_ThB\_\_2    & \textcolor{black}{25.098}          \\
                                                &                                       & VirtualAlloc\_large\_M\_\_2      & \textcolor{red}{19.466}            & VirtualAlloc\_large\_ThA\_\_2    & \textcolor{red}{18.634}            & VirtualAlloc\_large\_ThB\_\_2      & \textcolor{red}{18.749}            \\
        \end{tabular}}
    \label{tab:abs_ent_windows}
\end{table*}

\begin{figure}
    \centering
    \includegraphics[width=0.9\columnwidth]{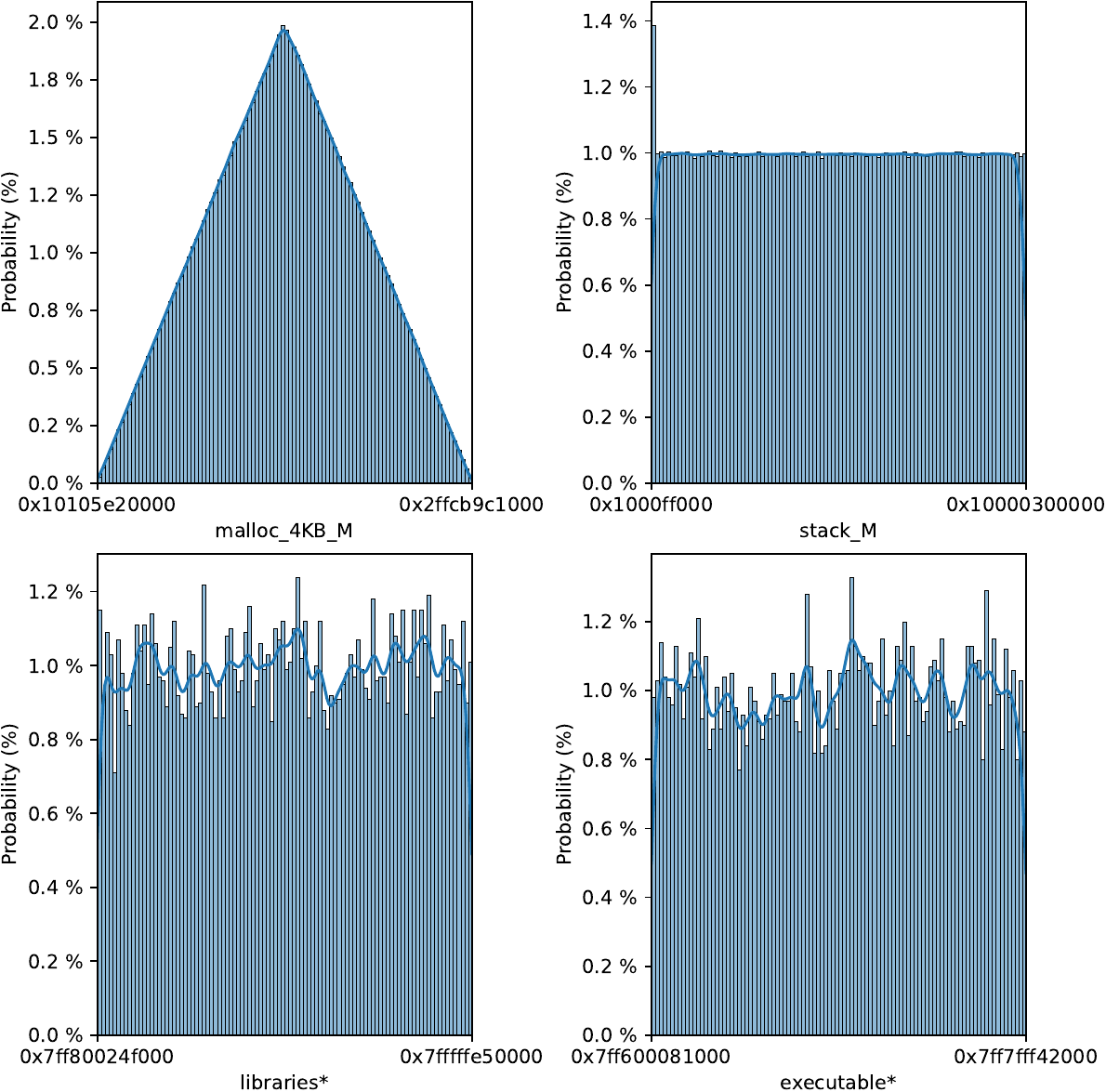}
    \caption{Probability Distribution in Windows 11. All the distributions are available at: \url{https://zenodo.org/records/12968870/files/windows_distribution.pdf}}
    \label{fig:distrib_win}
\end{figure}

\mypar{Correlation Entropy} In the Native MacOS, we have low Correlation Entropies due to the several objects allocated very close to each other. Although they are low, there is no need to have any prior knowledge since absolute entropies are as low as correlated ones, thus making bruteforce directly on the target object the preferred strategy. Instead, for Rosetta, we have four different memory regions, and we have higher Correlation Entropies. In particular, we have high Correlation Entropies between the objects of different regions. However, as in the Native MacOS system, the absolute entropies are so low that a direct bruteforce on the desired object is preferred.

\subsection{Windows 11}
\label{sec:windows}

Windows 11, like MacOS, has objects that are randomized at boot-time. These objects are the libraries and the executable. Hence, we collect 4,000,000 runtime samples and 20,000 boot-time samples. We compile our sample script with the \textit{MSVC} compiler, employing the flags \texttt{/DINAMICBASE} and \texttt{/HIGHENTROPYVA}. At the kernel level, we rely on the default \ac{ASLR} configurations, which are \texttt{High-entropy \ac{ASLR}} and \texttt{Bottom-up \ac{ASLR}}. These settings are intended to provide the highest level of \ac{ASLR} effectiveness in Windows.

\mypar{Allocation Layout} Looking at the allocation layout in Figure~\ref{fig:win_layout} we clearly identify two regions. The first one is located at low addresses and contains all the \texttt{malloc()} and \texttt{VirtualAlloc()} objects, as well as threads' stacks. At the other end of the memory, we can see the executable and the libraries, which are the objects whose position is determined at boot-time.

\mypar{Probability Distribution} Figure~\ref{fig:distrib_win} shows probability distributions for Windows. Here, we can observe three different shapes for the distribution. We have threads and stacks that are randomized almost uniformly; the lowest addresses are more likely to be used. The executable and libraries are also randomized uniformly, but given the lower amount of sample (they are randomized at boot-time), this is not clear from Figure~\ref{fig:distrib_win}. Finally, we have \texttt{malloc()} and \texttt{VirtualAlloc()} objects that are randomized following a triangular distribution. This last aspect usually means that the position is obtained by combining two independent sources of entropy, obtaining an Irwin-Hall distribution. This choice, even if it provides a larger entropy than the single random variable, potentially exposes the system to attacks regarding the most common value, reducing the absolute effort needed to de-randomize the section.

\mypar{Absolute Entropy} From Table~\ref{tab:abs_ent_windows}, we see an \AbsEntropy\ greater than 23 bits overall, with some objects reaching 31 bits. As expected, Large pages have less entropy compared to other sections due to a higher page alignment. However, we have low absolute entropies for boot-time randomized objects, i.e., the executable and the libraries. This poses a major risk to the security of the system since these objects are the most used for control-flow hijacking. Hence, bruteforce attacks are feasible in a short period of time.

\mypar{Correlation Entropy} Simirarly to the other \acp{OS}, we have what expected from the allocation layout: A high correlation inside the identified regions and a low correlation between regions. However, no relevant correlations emerge as the absolute entropies of relevant objects are lower than correlated ones.

\section{Practical Attacks to ASLR}
\label{sec:attacks_to_aslr}

While it is straightforward to defeat \ac{ASLR} when there are memory leaks, the practicality of attacks without direct memory leaks must be discussed. We consider the following attacks: \ding{192} \emph{bruteforce}, \ding{193} \emph{spraying attacks}, \ding{194} \emph{crossections}, and \ding{195} \emph{partial overwrite} attacks. 

For these attacks, we mainly consider CVEs and security articles related to Linux for several reasons. Among all the \acp{OS} considered, Linux is the only completely open-source, which facilitates the development of open-source software. This aids the discovery and analysis of vulnerabilities, as security analysts and researchers can perform analyses starting from the source code and can rely on state-of-the-art security tools developed by the community. Additionally, most commonly used applications and libraries on Linux are well-documented, which favors the development of PoCs and the creation of security articles, making them easier to follow and understand. Finally, Linux is the \ac{OS} with the most disclosed vulnerabilities~\cite{cves_per_product}. Hence, we can find more examples of attacks and vulnerabilities to analyze.

As memory corruptions happen in all the \acp{OS}, the attacks we discuss in this section are not exclusive to Linux, but they can be applied to other \acp{OS} as well. However, the feasibility of the exploitation must also consider all the other mitigations in place in the \ac{OS}. Nonetheless, if they are not sufficient to prevent exploitation, our findings in \ac{ASLR} implementations highlight weaknesses that must be addressed. 

In the following attacks, we consider an attacker capable of performing 300 tps as discussed in Section~\ref{subsec:threat_model}. We remind the reader that this value is realistic for both local and remote scenarios since an attacker can parallelize over several threads/processes and connections, respectively.

\label{sec:poc_folios}
\mypar{Bruteforce Attacks} This is the most naive attack to \ac{ASLR}. The attacker guesses the position of a memory object and executes the exploit with hardcoded positions. This is the primary threat that ASLR is designed to protect.  
Depending on the entropy of the memory object, the number of attempts to guess the position of the object varies. We argue that the entropy of the memory object in modern systems is not high enough, making this type of attack feasible in practice.
To prove our point, we demonstrate how the lower entropy generated by Linux Folios can be exploited in recent vulnerabilities (CVE-2021-3156)~\cite{sudoedit_cve}\footnote{The same vulnerability also affects MacOS Big Sur 11.2, MacOS Catalina 10.15.7, and MacOS Mojave 10.14.6~\cite{sudoedit_disclosure_apple}} to perform an attack in a reasonable amount of time (i.e., minutes). 
In particular, this CVE reports a heap overflow vulnerability on \texttt{sudo}, from version \texttt{1.8.2} (included) to \texttt{1.9.5p2} (excluded). Therefore, we set up an Ubuntu 22.04 environment, with kernel version 6.4.9 and \texttt{sudo 1.9.4}. In this setup, the glibc is allocated through Linux Folios. Hence, its memory page has a randomization entropy of only 19 bits.

We develop a Proof of Concept (PoC)\footnote{https://zenodo.org/doi/10.5281/zenodo.12784286} such that the overflow modifies a function pointer being called later in the execution. The new function pointer is a hardcoded glibc address that contains a gadget which performs a stack pivoting attack, moving the stack pointer close to the environment variables, where we place a ROP chain to achieve privilege escalation. 
The success rate of the exploit is directly influenced by the effectiveness of ASLR in randomizing glibc position. A robust ASLR implementation would render our exploit ineffective. However, the reality is that ASLR randomization does not entirely prevent the exploit's success. 

To measure the average time taken and to verify the entropy value, we run the exploit 500 times. Moreover, given that an attacker can also parallelize the exploits, we run 8, 4, and 2 exploits in parallel. Theoretically, with an entropy of 19 bits, the average number of attempts to guess the position of the glibc is $2^{19} = 524,288$ attempts. As shown in Table~\ref{table:poc_results}, we achieve an average tps of 283 and an average number of attempts of $555,765$, which is very close to the expected value. Moreover, we can see that even with one exploit, the time to achieve privilege escalation is 32 minutes and 44 seconds, which is very low. It is even lower when 8 exploits are run in parallel. In this case, the first exploit achieving privilege escalation takes only 4 minutes with an average of $58,776$ attempts. In addition, all the exploits running in parallel generate an average tps of 1,904, underlining the possibility of achieving a higher number of tps w.r.t. the reference value of 300 tps considered in Section~\ref{subsec:threat_model}.

This PoC highlights the problems arising from low \AbsEntropy : relevant memory objects like the libraries and the executable should have a high \AbsEntropy\ since these regions are very often needed by attackers to obtain command execution. Therefore, the low \AbsEntropy\ of the glibc in Linux Folios is a significant security risk. We believe that the entropy of relevant objects should be increased to a minimum of 24 bits (i.e., $2^{24} = 16,777,216$) which can be bruteforeced in more than a day with 300 tps on average. Instead, we consider a good entropy value to be 30+ bits, which would require more than a month to be bruteforced.

\begin{table}
    \centering
    \caption{Average attempts and time for privilege escalation.}
    \small
    \begin{tabular}{|c|c|c|c|}
        \hline
        \textbf{\# Parallel Exploits} & \textbf{Avg Attempts} & \textbf{Avg Time} & \textbf{Avg TPS} \\
        \hline
        8 & 58,776 & 00:04:06::904 & 1,904 \\
        \hline
        4 & 128,884 & 00:08:23::539 & 1,023 \\
        \hline
        2 & 275,585 & 00:16:51::781 & 544 \\
        \hline
        \textbf{1} &  \textbf{555,765} & \textbf{00:32:44::597} & \textbf{283} \\
        \hline
    \end{tabular}
    \label{table:poc_results}
\end{table}

\mypar{Spraying Attacks} Even when the absolute entropy is high (>27 bits), \textit{Spraying Attacks} can drastically reduce it by repeating the same data for several memory pages. This attack aims to repeat the same payload across several memory pages to increase the likelihood of successfully referencing the pointer.

Recently, a vulnerability on the glibc (CVE-2023-4911, a.k.a. Looney Tunables) has been exploited using a stack spraying technique to achieve local privilege escalation~\cite{looney_tunables}. The attack exploits a buffer overflow vulnerability on the Linux dynamic loader \texttt{ld.so} to change the string pointer of \texttt{l\_info[DT\_RPATH]}, a string that specifies the libraries search path. In particular, the attack modifies the string pointer to point to the environment variables, where a string representing an attacker-controlled directory resides. Under Linux, the environment variables have 24.5 bits of entropy. This value does not guarantee strong randomization since it requires an average of $2^{24.5} = 23,726,566$ attempts to locate the precise address of the environment variables. However, repeating the same string across several memory pages increases the likelihood of successfully referencing the pointer. Given that the maximum size for environment variables in Linux is \texttt{6MB}, and if the same string spans multiple pages, the likelihood of successful exploitation improves to 1 over 4,096 (i.e., $\frac{24GB}{6MB}$). This means that, at a rate of 300 tps, local privilege escalation is achieved in about 14 seconds.

Spraying attacks are feasible in any memory region an attacker can control exhaustively. For instance, other objects that can be used for spraying attacks are all the dynamically allocated pages, such as the heap. In Linux, the heap has an entropy of 28.8 bits, which is the maximum entropy in the system. Nonetheless, this may not deter an attacker capable of allocating gigabytes of memory pages without a limit on heap growth. To make these attacks less relibale, user-controlled memory pages should have a higher absolute entropy (>30 bits).

\mypar{Crossections Attacks}
When the entropy is high enough, bruteforcing a precise memory location may become unrealistic. However, if there are \textit{memory leak} vulnerabilities, the attacker may rely on this information to exploit the correlation between memory objects~\cite{aslr_bypass_ios,aslr_bypass_linux}. As we pointed out in Section~\ref{sec:results}, strong correlation paths exist between objects that are closely allocated in memory. For instance, the executable and heap entropy in Linux are 28.8 bits. Hence, the average number of attempts to bruteforce a precise address of one of the two objects is $2^{28.8} = 467,373,275$ attempts, or $1,557,911$ seconds ($18$ days) if we consider 300 tps. However, with a memory leak on the heap, an attacker can bruteforce a precise address of the executable in $2^{13} = 8,192$ attempts, on average, which is approximately $27$ seconds. 

Although this example highlights the risks of having a strong \CorrEntropy\ between two memory objects, this is less relevant than having a low absolute entropy; a weak \CorrEntropy\ between two memory objects is useless if the two objects have weak Absolute Entropies as well. This happens, for instance, on MacOS.

\mypar{Partial Overwrite Attacks}
In some situations, an attacker can partially overwrite a function pointer to hijack the control flow, potentially compromising the security of a system. For instance, by changing the lower bits of a function pointer it is possible to call a different function within the same library, or a function that belongs to relatively close, or contiguous, libraries. For this reason, a good \ac{ASLR} implementation should allocate libraries with a very high \CorrEntropy\ one another. This is not the case for Linux, where, during program startup, all the libraries are allocated contiguously or with low entropy to one another. Therefore, whenever a vulnerability allows a partial overwrite of a function pointer, the difference between the original and the targeted one is always the same. However, \ac{ASLR} changes most of the bits of the addresses and thus, the chances of correctly modifying the lower bits of the address depend on the distance between the two pointers and the page alignment. For instance, with a page alignment of \texttt{4KB} and a pointers difference of less than 256 bytes, the attack is completely deterministic. Instead, with a pointers difference lower than a standard page (\texttt{4KB}), the attack requires an average of 16 attempts (4 bits of randomness). Finally, with a pointers difference higher than a standard page, the attack requires an average of 4,096 attempts (12 bits of randomness). However, with Linux folios, libraries with a size larger than \texttt{2MB}, such as the \texttt{glibc}, have a page alignment of \texttt{2MB}. Therefore, whenever the pointer difference is lower than \texttt{64KB}, the attack may be completely deterministic; if the targeted pointer belongs to the same library as the original one, the attack is completely deterministic. Otherwise, there might be at least 9 bits of entropy due to the strong correlation between libraries allocated with Folio and standard pages, as we can see in Figure~\ref{fig:corr_lin649}.

This attack should highlight two major problems in \ac{ASLR} implementations: \ding{192} libraries should never be contiguous with one another, but instead, they should be independent (high \CorrEntropy), and \ding{193} higher page alignments drastically reduce the chance of exploitation.

\section{Long Life to ASLR}
\label{sec:long-life-aslr}

As we mentioned in Section~\ref{sec:results}, modern \ac{ASLR} implementations have several flaws. MacOS is characterized by a low absolute entropy overall, with entropy for relevant sections such as the executable and the libraries ranging from 11.5 to 15.5 bits. Windows 11 has better peaks, reaching up to 31 bits of entropy, but it also has a poor randomization where it matters the most: the executable and the libraries, reaching at most 19 bits of entropy. Instead, Linux achieves high entropies in all the relevant objects, including the stack and the heap, resulting in the best \ac{OS} in randomization. However, the newest kernel versions and memory leaks can considerably reduce the effort of bruteforce attacks.
Ideally, any memory object in the system should reach an absolute entropy of 31 bits, such as one of the small allocated objects in Windows 11, to consider the system resilient against bruteforce attacks. If we consider 300 tps and an average of 2 billion attempts ($2^{31}$), it would require an average of approximately 83 days to guess the position of an object in memory, which is not practical, especially if Intrusion Detection Systems (IDSs) are in place.

Therefore, our first proposal consists of a virtual memory fragmentation where each relevant object has its own region. For instance, in Linux, the addresses below the executable and the heap are not used. Hence, there is a place for objects, like the executable or the heap, also to resolve the correlation between these two objects. Similarly, libraries can be moved to lower addresses to remove the correlation between this latter and allocated pages.

Secondly, we propose the utilization of 5-level page tables so that the available bits in an address extend from 48 to 57. We already have processors that support 5-level page tables, but modern systems seem not to adopt them. The reason is probably related to performance once more; one memory access to virtual memory results in six accesses to physical memory instead of five, thus slightly increasing the memory access time. 


\section{Conclusions}
\label{sec:conclusions}

In this paper, we conduct the first comprehensive evaluation of \ac{ASLR} effectiveness across major platforms through the statistical analysis of memory object positions. Adopting a low-bias estimator, the NSB estimator, allows us to reduce the necessary sample size, thereby facilitating the analysis of reboot randomization, a notably time-intensive operation.

We highlight significant weaknesses of current implementations of ASLR, like the lack of entropy of libraries and executable objects in Windows and MacOS or the entropy reduction found in recent Linux distributions (introduced with Folios). Overall, Linux distributions provide the best randomization- still insufficient to stand against modern exploitation techniques - while Windows and MacOS fail to randomize key memory areas like executable code and libraries adequately. Our findings highlight opportunities for OS vendors to strengthen implementations and better protect users from malicious attacks. Addressing reduced entropy from correlations, optimizing allocation patterns, and increasing object granularity could all fortify defenses. Moreover, this research suggests that the evolution of operating systems is often not security-focused, and the introduction of Linux Folios confirms this claim. We expect major changes with the broad adoption of a 5-level paging system, providing by construction more bits to the randomization process \cite{mmanagement_linux5.15}. Another aspect highlighted in this work is the role of allocation patterns and the difficulty of correctly modeling such behavior. Real-world software is a complex ecosystem of interacting objects, and their performance may vary significantly from expectations and often be lower.

\section{Acknowledgements}
Lorenzo Binosi acknowledges support from TIM S.p.A. through the PhD scholarship. Additionally, this work was partially supported by Project FARE (PNRR M4.C2.1.1 PRIN 2022, Cod. 202225BZJC, CUP D53D23008380006, Avviso D.D 104 02.02.2022) and Project SETA (PNRR M4.C2.1.1 PRIN 2022 PNRR, Cod. P202233M9Z, CUP F53D23009120001, Avviso D.D 1409 14.09.2022). Both projects are under the Italian NRRP MUR program funded by the European Union - NextGenerationEU. Finally, this work was partially supported by project SERICS (PE00000014) under the MUR National Recovery and Resilience Plan funded by the European Union - NextGenerationEU.

\bibliographystyle{ACM-Reference-Format}
\bibliography{paper.bib}


\begin{thebibliography}{45}


\ifx \showCODEN    \undefined \def \showCODEN     #1{\unskip}     \fi
\ifx \showDOI      \undefined \def \showDOI       #1{#1}\fi
\ifx \showISBNx    \undefined \def \showISBNx     #1{\unskip}     \fi
\ifx \showISBNxiii \undefined \def \showISBNxiii  #1{\unskip}     \fi
\ifx \showISSN     \undefined \def \showISSN      #1{\unskip}     \fi
\ifx \showLCCN     \undefined \def \showLCCN      #1{\unskip}     \fi
\ifx \shownote     \undefined \def \shownote      #1{#1}          \fi
\ifx \showarticletitle \undefined \def \showarticletitle #1{#1}   \fi
\ifx \showURL      \undefined \def \showURL       {\relax}        \fi
\providecommand\bibfield[2]{#2}
\providecommand\bibinfo[2]{#2}
\providecommand\natexlab[1]{#1}
\providecommand\showeprint[2][]{arXiv:#2}

\bibitem[Acharya et~al\mbox{.}(2017)]%
        {DBLP:journals/tit/AcharyaOST17}
\bibfield{author}{\bibinfo{person}{Jayadev Acharya}, \bibinfo{person}{Alon
  Orlitsky}, \bibinfo{person}{Ananda~Theertha Suresh}, {and}
  \bibinfo{person}{Himanshu Tyagi}.} \bibinfo{year}{2017}\natexlab{}.
\newblock \showarticletitle{Estimating Renyi Entropy of Discrete
  Distributions}.
\newblock \bibinfo{journal}{\emph{{IEEE} Trans. Inf. Theory}}
  \bibinfo{volume}{63}, \bibinfo{number}{1} (\bibinfo{year}{2017}),
  \bibinfo{pages}{38--56}.
\newblock
\urldef\tempurl%
\url{https://doi.org/10.1109/TIT.2016.2620435}
\showDOI{\tempurl}


\bibitem[Advisory(2023)]%
        {looney_tunables}
\bibfield{author}{\bibinfo{person}{Qualys~Security Advisory}.}
  \bibinfo{year}{2023}\natexlab{}.
\newblock \showarticletitle{Looney Tunables: Local Privilege Escalation in the
  glibc's ld.so (CVE-2023-4911)}.
\newblock \bibinfo{journal}{\emph{qualys.com}} (\bibinfo{date}{Oct}
  \bibinfo{year}{2023}).
\newblock
\urldef\tempurl%
\url{https://www.qualys.com/2023/10/03/cve-2023-4911/looney-tunables-local-privilege-escalation-glibc-ld-so.txt}
\showURL{%
\tempurl}


\bibitem[Android(2024a)]%
        {android_mmu_configurations}
\bibfield{author}{\bibinfo{person}{Android}.} \bibinfo{year}{2024}\natexlab{a}.
\newblock \showarticletitle{Android Memory Management Unit Configurations}.
\newblock \bibinfo{journal}{\emph{cs.android.com}} (\bibinfo{date}{Jul}
  \bibinfo{year}{2024}).
\newblock
\urldef\tempurl%
\url{https://cs.android.com/android/platform/superproject/main/+/main:external/trusty/lk/arch/arm64/include/arch/arm64/mmu.h}
\showURL{%
\tempurl}


\bibitem[Android(2024b)]%
        {android_scudo}
\bibfield{author}{\bibinfo{person}{Android}.} \bibinfo{year}{2024}\natexlab{b}.
\newblock \showarticletitle{Android Scudo Heap Allocator}.
\newblock \bibinfo{journal}{\emph{source.android.com}} (\bibinfo{date}{Apr}
  \bibinfo{year}{2024}).
\newblock
\urldef\tempurl%
\url{https://source.android.com/docs/security/test/scudo}
\showURL{%
\tempurl}


\bibitem[Android(2024c)]%
        {android_studio}
\bibfield{author}{\bibinfo{person}{Android}.} \bibinfo{year}{2024}\natexlab{c}.
\newblock \showarticletitle{Android Studio}.
\newblock \bibinfo{journal}{\emph{developer.android.com}} (\bibinfo{date}{Jul}
  \bibinfo{year}{2024}).
\newblock
\urldef\tempurl%
\url{https://developer.android.com/studio}
\showURL{%
\tempurl}


\bibitem[Apple(2021)]%
        {sudoedit_disclosure_apple}
\bibfield{author}{\bibinfo{person}{Apple}.} \bibinfo{year}{2021}\natexlab{}.
\newblock \showarticletitle{About the security content of macOS Big Sur 11.2.1,
  macOS Catalina 10.15.7 Supplemental Update, and macOS Mojave 10.14.6}.
\newblock \bibinfo{journal}{\emph{support.apple.com}} (\bibinfo{date}{Feb}
  \bibinfo{year}{2021}).
\newblock
\urldef\tempurl%
\url{https://support.apple.com/en-ca/103058}
\showURL{%
\tempurl}


\bibitem[Apple(2024)]%
        {apple_arm_translation_code}
\bibfield{author}{\bibinfo{person}{Apple}.} \bibinfo{year}{2024}\natexlab{}.
\newblock \showarticletitle{MacOS Virtual Address Translation}.
\newblock \bibinfo{journal}{\emph{opensource.apple.com}} (\bibinfo{date}{Jul}
  \bibinfo{year}{2024}).
\newblock
\urldef\tempurl%
\url{https://opensource.apple.com/source/xnu/xnu-7195.81.3/osfmk/arm64/proc_reg.h.auto.html}
\showURL{%
\tempurl}


\bibitem[Aristizabal et~al\mbox{.}(2013)]%
        {aristizabal2013measuring}
\bibfield{author}{\bibinfo{person}{David~Herrera Aristizabal},
  \bibinfo{person}{David~Mora Rodriguez}, {and} \bibinfo{person}{Ricardo~Yepes
  Guevara}.} \bibinfo{year}{2013}\natexlab{}.
\newblock \showarticletitle{Measuring ASLR implementations on modern operating
  systems}. In \bibinfo{booktitle}{\emph{2013 47th International Carnahan
  Conference on Security Technology (ICCST)}}. IEEE, \bibinfo{pages}{1--6}.
\newblock


\bibitem[ARM(2019)]%
        {armv8a_address_translation}
\bibfield{author}{\bibinfo{person}{ARM}.} \bibinfo{year}{2019}\natexlab{}.
\newblock \showarticletitle{ARMv8-A Address Translation}.
\newblock \bibinfo{journal}{\emph{documentation-service.arm.com}}
  (\bibinfo{date}{Jul} \bibinfo{year}{2019}).
\newblock
\urldef\tempurl%
\url{https://documentation-service.arm.com/static/5efa1d23dbdee951c1ccdec5}
\showURL{%
\tempurl}


\bibitem[CVEdetails.com(2024)]%
        {cves_per_product}
\bibfield{author}{\bibinfo{person}{CVEdetails.com}.}
  \bibinfo{year}{2024}\natexlab{}.
\newblock \showarticletitle{Top 50 Products By Total Number Of Distinct
  Vulnerabilities}.
\newblock \bibinfo{journal}{\emph{cvedetails.com}} (\bibinfo{date}{Jul}
  \bibinfo{year}{2024}).
\newblock
\urldef\tempurl%
\url{https://www.cvedetails.com/top-50-products.php}
\showURL{%
\tempurl}


\bibitem[D{\'\i}az et~al\mbox{.}(2021)]%
        {diaz2021address}
\bibfield{author}{\bibinfo{person}{Raquel~V{\'a}zquez D{\'\i}az},
  \bibinfo{person}{Marti{\~n}o Rivera-Dourado}, \bibinfo{person}{Rub{\'e}n
  P{\'e}rez-Jove}, \bibinfo{person}{Pilar~Vila Avenda{\~n}o}, {and}
  \bibinfo{person}{Jos{\'e}~M V{\'a}zquez-Naya}.}
  \bibinfo{year}{2021}\natexlab{}.
\newblock \showarticletitle{Address Space Layout Randomization Comparative
  Analysis on Windows 10 and Ubuntu 18.04 LTS}.
\newblock \bibinfo{journal}{\emph{Engineering Proceedings}}
  \bibinfo{volume}{7}, \bibinfo{number}{1} (\bibinfo{year}{2021}),
  \bibinfo{pages}{26}.
\newblock


\bibitem[Ding et~al\mbox{.}(2014)]%
        {DBLP:conf/icc/DingPZZ14}
\bibfield{author}{\bibinfo{person}{Yu Ding}, \bibinfo{person}{Zhuo Peng},
  \bibinfo{person}{Yuanyuan Zhou}, {and} \bibinfo{person}{Chao Zhang}.}
  \bibinfo{year}{2014}\natexlab{}.
\newblock \showarticletitle{Android low entropy demystified}. In
  \bibinfo{booktitle}{\emph{{IEEE} International Conference on Communications,
  {ICC} 2014, Sydney, Australia, June 10-14, 2014}}.
  \bibinfo{publisher}{{IEEE}}, \bibinfo{pages}{659--664}.
\newblock
\urldef\tempurl%
\url{https://doi.org/10.1109/ICC.2014.6883394}
\showDOI{\tempurl}


\bibitem[Díaz et~al\mbox{.}(2019)]%
        {aslr_tool}
\bibfield{author}{\bibinfo{person}{Raquel~Vázquez Díaz},
  \bibinfo{person}{Martiño Rivera-Dourado}, \bibinfo{person}{Rubén
  Pérez-Jove}, \bibinfo{person}{Pilar~Vila Avendaño}, {and}
  \bibinfo{person}{José~M. Vázquez-Naya}.} \bibinfo{year}{2019}\natexlab{}.
\newblock \showarticletitle{ASLR analyzer}.
\newblock \bibinfo{journal}{\emph{GitHub}} (\bibinfo{year}{2019}).
\newblock
\urldef\tempurl%
\url{https://github.com/raquelvqz/aslr/}
\showURL{%
\tempurl}


\bibitem[Groß(2021)]%
        {aslr_bypass_linux}
\bibfield{author}{\bibinfo{person}{Samuel Groß}.}
  \bibinfo{year}{2021}\natexlab{}.
\newblock \showarticletitle{Breaking 64 bit aslr on Linux x86-64}.
\newblock \bibinfo{journal}{\emph{googleprojectzero.blogspot.com}}
  (\bibinfo{date}{Jan} \bibinfo{year}{2021}).
\newblock
\urldef\tempurl%
\url{https://googleprojectzero.blogspot.com/2020/01/remote-iphone-exploitation-part-2.html}
\showURL{%
\tempurl}


\bibitem[Herlands et~al\mbox{.}(2014)]%
        {DBLP:conf/uss/HerlandsHD14}
\bibfield{author}{\bibinfo{person}{William Herlands}, \bibinfo{person}{Thomas
  Hobson}, {and} \bibinfo{person}{Paula~J. Donovan}.}
  \bibinfo{year}{2014}\natexlab{}.
\newblock \showarticletitle{Effective Entropy: Security-Centric Metric for
  Memory Randomization Techniques}. In \bibinfo{booktitle}{\emph{7th Workshop
  on Cyber Security Experimentation and Test, {CSET} '14, San Diego, CA, USA,
  August 18, 2014}}, \bibfield{editor}{\bibinfo{person}{Chris Kanich} {and}
  \bibinfo{person}{Patrick Lardieri}} (Eds.). \bibinfo{publisher}{{USENIX}
  Association}.
\newblock
\urldef\tempurl%
\url{https://www.usenix.org/conference/cset14/workshop-program/presentation/herlands}
\showURL{%
\tempurl}


\bibitem[Hern{\'{a}}ndez and Samengo(2019)]%
        {DBLP:journals/entropy/HernandezS19}
\bibfield{author}{\bibinfo{person}{Dami{\'{a}}n~G. Hern{\'{a}}ndez} {and}
  \bibinfo{person}{In{\'{e}}s Samengo}.} \bibinfo{year}{2019}\natexlab{}.
\newblock \showarticletitle{Estimating the Mutual Information between Two
  Discrete, Asymmetric Variables with Limited Samples}.
\newblock \bibinfo{journal}{\emph{Entropy}} \bibinfo{volume}{21},
  \bibinfo{number}{6} (\bibinfo{year}{2019}), \bibinfo{pages}{623}.
\newblock
\urldef\tempurl%
\url{https://doi.org/10.3390/E21060623}
\showDOI{\tempurl}


\bibitem[Kaplan et~al\mbox{.}(2014)]%
        {DBLP:conf/woot/KaplanKHD14}
\bibfield{author}{\bibinfo{person}{David Kaplan}, \bibinfo{person}{Sagi Kedmi},
  \bibinfo{person}{Roee Hay}, {and} \bibinfo{person}{Avi Dayan}.}
  \bibinfo{year}{2014}\natexlab{}.
\newblock \showarticletitle{Attacking the Linux {PRNG} On Android: Weaknesses
  in Seeding of Entropic Pools and Low Boot-Time Entropy}. In
  \bibinfo{booktitle}{\emph{8th {USENIX} Workshop on Offensive Technologies,
  {WOOT} '14, San Diego, CA, USA, August 19, 2014}},
  \bibfield{editor}{\bibinfo{person}{Sergey Bratus} {and}
  \bibinfo{person}{Felix~"FX" Lindner}} (Eds.). \bibinfo{publisher}{{USENIX}
  Association}.
\newblock
\urldef\tempurl%
\url{https://www.usenix.org/conference/woot14/workshop-program/presentation/kaplan}
\showURL{%
\tempurl}


\bibitem[kernel~development community(2023a)]%
        {GlibcMalloc}
\bibfield{author}{\bibinfo{person}{The kernel~development community}.}
  \bibinfo{year}{2023}\natexlab{a}.
\newblock \showarticletitle{LibC: Malloc Tunable Parameters}.
\newblock \bibinfo{journal}{\emph{gnu.org}} (\bibinfo{date}{Jan}
  \bibinfo{year}{2023}).
\newblock
\urldef\tempurl%
\url{https://www.gnu.org/software/libc/manual/2.37/html_mono/libc.html\#Malloc-Tunable-Parameters}
\showURL{%
\tempurl}


\bibitem[kernel~development community(2023b)]%
        {linux_folios}
\bibfield{author}{\bibinfo{person}{The kernel~development community}.}
  \bibinfo{year}{2023}\natexlab{b}.
\newblock \showarticletitle{Memory Management APIs Documentation}.
\newblock \bibinfo{journal}{\emph{kernel.org}} (\bibinfo{date}{Jan}
  \bibinfo{year}{2023}).
\newblock
\urldef\tempurl%
\url{https://www.kernel.org/doc/html/v6.0/core-api/mm-api.html}
\showURL{%
\tempurl}


\bibitem[Kerrisk(2023)]%
        {pthread_create}
\bibfield{author}{\bibinfo{person}{Michael Kerrisk}.}
  \bibinfo{year}{2023}\natexlab{}.
\newblock \showarticletitle{pthread\_create(3) - Linux Manual}.
\newblock \bibinfo{journal}{\emph{man7.org}} (\bibinfo{date}{Jan}
  \bibinfo{year}{2023}).
\newblock
\urldef\tempurl%
\url{https://man7.org/linux/man-pages/man3/pthread_create.3.html}
\showURL{%
\tempurl}


\bibitem[Lee et~al\mbox{.}(2014)]%
        {DBLP:conf/sp/LeeLWKL14}
\bibfield{author}{\bibinfo{person}{Byoungyoung Lee}, \bibinfo{person}{Long Lu},
  \bibinfo{person}{Tielei Wang}, \bibinfo{person}{Taesoo Kim}, {and}
  \bibinfo{person}{Wenke Lee}.} \bibinfo{year}{2014}\natexlab{}.
\newblock \showarticletitle{From Zygote to Morula: Fortifying Weakened {ASLR}
  on Android}. In \bibinfo{booktitle}{\emph{2014 {IEEE} Symposium on Security
  and Privacy, {SP} 2014, Berkeley, CA, USA, May 18-21, 2014}}.
  \bibinfo{publisher}{{IEEE} Computer Society}, \bibinfo{pages}{424--439}.
\newblock
\urldef\tempurl%
\url{https://doi.org/10.1109/SP.2014.34}
\showDOI{\tempurl}


\bibitem[Li et~al\mbox{.}(2006)]%
        {DBLP:conf/acsac/LiJS06}
\bibfield{author}{\bibinfo{person}{Lixin Li}, \bibinfo{person}{James~E. Just},
  {and} \bibinfo{person}{R. Sekar}.} \bibinfo{year}{2006}\natexlab{}.
\newblock \showarticletitle{Address-Space Randomization for Windows Systems}.
  In \bibinfo{booktitle}{\emph{22nd Annual Computer Security Applications
  Conference {(ACSAC} 2006), 11-15 December 2006, Miami Beach, Florida,
  {USA}}}. \bibinfo{publisher}{{IEEE} Computer Society},
  \bibinfo{pages}{329--338}.
\newblock
\urldef\tempurl%
\url{https://doi.org/10.1109/ACSAC.2006.10}
\showDOI{\tempurl}


\bibitem[Liebergeld and Lange(2013)]%
        {DBLP:conf/iscis/LiebergeldL13}
\bibfield{author}{\bibinfo{person}{Steffen Liebergeld} {and}
  \bibinfo{person}{Matthias Lange}.} \bibinfo{year}{2013}\natexlab{}.
\newblock \showarticletitle{Android Security, Pitfalls and Lessons Learned}. In
  \bibinfo{booktitle}{\emph{Information Sciences and Systems 2013 - Proceedings
  of the 28th International Symposium on Computer and Information Sciences,
  {ISCIS} 2013, Paris, France, October 28-29, 2013}}
  \emph{(\bibinfo{series}{Lecture Notes in Electrical Engineering},
  Vol.~\bibinfo{volume}{264})}, \bibfield{editor}{\bibinfo{person}{Erol
  Gelenbe} {and} \bibinfo{person}{Ricardo Lent}} (Eds.).
  \bibinfo{publisher}{Springer}, \bibinfo{pages}{409--417}.
\newblock
\urldef\tempurl%
\url{https://doi.org/10.1007/978-3-319-01604-7\_40}
\showDOI{\tempurl}


\bibitem[Marco-Gisbert and Ripoll(2014)]%
        {marco2014effectiveness}
\bibfield{author}{\bibinfo{person}{Hector Marco-Gisbert} {and}
  \bibinfo{person}{Ismael Ripoll}.} \bibinfo{year}{2014}\natexlab{}.
\newblock \showarticletitle{On the Effectiveness of Full-ASLR on 64-bit Linux}.
  In \bibinfo{booktitle}{\emph{Proceedings of the In-Depth Security
  Conference}}.
\newblock


\bibitem[{Marco Gisbert} and Ripoll(2016a)]%
        {ASLRA_tool}
\bibfield{author}{\bibinfo{person}{Hector {Marco Gisbert}} {and}
  \bibinfo{person}{Ismael Ripoll}.} \bibinfo{year}{2016}\natexlab{a}.
\newblock \showarticletitle{ASLRA - ASLR Analyzer}.
\newblock \bibinfo{journal}{\emph{cybersecurity.upv.es}} (\bibinfo{date}{Apr}
  \bibinfo{year}{2016}).
\newblock
\urldef\tempurl%
\url{https://web.archive.org/web/20230306172439/http://cybersecurity.upv.es/tools/aslra/aslr-analyzer.html}
\showURL{%
\tempurl}


\bibitem[{Marco Gisbert} and Ripoll(2016b)]%
        {Gisbert2016ExploitingLA_PositiveEnt}
\bibfield{author}{\bibinfo{person}{Hector {Marco Gisbert}} {and}
  \bibinfo{person}{Ismael Ripoll}.} \bibinfo{year}{2016}\natexlab{b}.
\newblock \showarticletitle{Exploiting Linux and PaX ASLR{\textquoteright}s
  weaknesses on 32-bit and 64-bit systems}.
\newblock \bibinfo{journal}{\emph{BlackHat Asia}} (\bibinfo{date}{29 March}
  \bibinfo{year}{2016}).
\newblock


\bibitem[Marco-Gisbert and Ripoll~Ripoll(2019)]%
        {marco2019address}
\bibfield{author}{\bibinfo{person}{Hector Marco-Gisbert} {and}
  \bibinfo{person}{Ismael Ripoll~Ripoll}.} \bibinfo{year}{2019}\natexlab{}.
\newblock \showarticletitle{Address space layout randomization next
  generation}.
\newblock \bibinfo{journal}{\emph{Applied Sciences}} \bibinfo{volume}{9},
  \bibinfo{number}{14} (\bibinfo{year}{2019}), \bibinfo{pages}{2928}.
\newblock


\bibitem[Nemenman(2012)]%
        {nsb_website}
\bibfield{author}{\bibinfo{person}{Ilya Nemenman}.}
  \bibinfo{year}{2012}\natexlab{}.
\newblock \showarticletitle{Nemenman-Shafee-Bialek (NSB) Entropy Estimator}.
\newblock \bibinfo{journal}{\emph{nemenmanlab.org}} (\bibinfo{year}{2012}).
\newblock
\urldef\tempurl%
\url{https://nemenmanlab.org/~ilya/index.php/Entropy_Estimation}
\showURL{%
\tempurl}


\bibitem[Nemenman et~al\mbox{.}(2001)]%
        {DBLP:conf/nips/NemenmanSB01}
\bibfield{author}{\bibinfo{person}{Ilya Nemenman}, \bibinfo{person}{F. Shafee},
  {and} \bibinfo{person}{William Bialek}.} \bibinfo{year}{2001}\natexlab{}.
\newblock \showarticletitle{Entropy and Inference, Revisited}. In
  \bibinfo{booktitle}{\emph{Advances in Neural Information Processing Systems
  14 [Neural Information Processing Systems: Natural and Synthetic, {NIPS}
  2001, December 3-8, 2001, Vancouver, British Columbia, Canada]}},
  \bibfield{editor}{\bibinfo{person}{Thomas~G. Dietterich},
  \bibinfo{person}{Suzanna Becker}, {and} \bibinfo{person}{Zoubin Ghahramani}}
  (Eds.). \bibinfo{publisher}{{MIT} Press}, \bibinfo{pages}{471--478}.
\newblock
\urldef\tempurl%
\url{https://proceedings.neurips.cc/paper/2001/hash/d46e1fcf4c07ce4a69ee07e4134bcef1-Abstract.html}
\showURL{%
\tempurl}


\bibitem[nick0ve(2021)]%
        {aslr_bypass_ios}
\bibfield{author}{\bibinfo{person}{nick0ve}.} \bibinfo{year}{2021}\natexlab{}.
\newblock \showarticletitle{Breaking 64 bit aslr on Linux x86-64}.
\newblock \bibinfo{journal}{\emph{github.com}} (\bibinfo{date}{Nov}
  \bibinfo{year}{2021}).
\newblock
\urldef\tempurl%
\url{https://github.com/nick0ve/how-to-bypass-aslr-on-linux-x86_64}
\showURL{%
\tempurl}


\bibitem[nvd.nist.gov(2021)]%
        {sudoedit_cve}
\bibfield{author}{\bibinfo{person}{nvd.nist.gov}.}
  \bibinfo{year}{2021}\natexlab{}.
\newblock \showarticletitle{Sudoedit CVE-2021-3156}.
\newblock \bibinfo{journal}{\emph{nvd.nist.gov}} (\bibinfo{date}{Jan}
  \bibinfo{year}{2021}).
\newblock
\urldef\tempurl%
\url{https://nvd.nist.gov/vuln/detail/CVE-2021-3156}
\showURL{%
\tempurl}


\bibitem[Scorecard(2023)]%
        {CVE_stats}
\bibfield{author}{\bibinfo{person}{Security Scorecard}.}
  \bibinfo{year}{2023}\natexlab{}.
\newblock \showarticletitle{CVE security vulnerability database. Security
  vulnerabilities, exploits.}
\newblock \bibinfo{journal}{\emph{CVE security vulnerability database. Security
  vulnerabilities, exploits, references and more.}} (\bibinfo{date}{Aug}
  \bibinfo{year}{2023}).
\newblock
\urldef\tempurl%
\url{https://www.cvedetails.com/vulnerabilities-by-types.php}
\showURL{%
\tempurl}


\bibitem[Shannon(1948)]%
        {DBLP:journals/bstj/Shannon48}
\bibfield{author}{\bibinfo{person}{Claude~E. Shannon}.}
  \bibinfo{year}{1948}\natexlab{}.
\newblock \showarticletitle{A mathematical theory of communication}.
\newblock \bibinfo{journal}{\emph{Bell Syst. Tech. J.}} \bibinfo{volume}{27},
  \bibinfo{number}{3} (\bibinfo{year}{1948}), \bibinfo{pages}{379--423}.
\newblock
\urldef\tempurl%
\url{https://doi.org/10.1002/J.1538-7305.1948.TB01338.X}
\showDOI{\tempurl}


\bibitem[Stats(2023a)]%
        {StatCounter_Desktop}
\bibfield{author}{\bibinfo{person}{Statcounter~Global Stats}.}
  \bibinfo{year}{2023}\natexlab{a}.
\newblock \showarticletitle{Desktop Operating System Market Share Worldwide}.
\newblock \bibinfo{journal}{\emph{StatCounter Global Stats}}
  (\bibinfo{date}{Aug} \bibinfo{year}{2023}).
\newblock
\urldef\tempurl%
\url{https://gs.statcounter.com/os-market-share/desktop/worldwide}
\showURL{%
\tempurl}


\bibitem[Stats(2023b)]%
        {StatCounter_Mobile}
\bibfield{author}{\bibinfo{person}{Statcounter~Global Stats}.}
  \bibinfo{year}{2023}\natexlab{b}.
\newblock \showarticletitle{Mobile Operating System Market Share Worldwide}.
\newblock \bibinfo{journal}{\emph{StatCounter Global Stats}}
  (\bibinfo{date}{Aug} \bibinfo{year}{2023}).
\newblock
\urldef\tempurl%
\url{https://gs.statcounter.com/os-market-share/mobile/worldwide}
\showURL{%
\tempurl}


\bibitem[Telegram(2024)]%
        {telegram_cves}
\bibfield{author}{\bibinfo{person}{Telegram}.} \bibinfo{year}{2024}\natexlab{}.
\newblock \showarticletitle{Telegram Vulnerabilities Archive}.
\newblock \bibinfo{journal}{\emph{opencve.io}} (\bibinfo{date}{Jun}
  \bibinfo{year}{2024}).
\newblock
\urldef\tempurl%
\url{https://www.opencve.io/cve?vendor=telegram&product=telegram}
\showURL{%
\tempurl}


\bibitem[Torvalds(2021a)]%
        {folios_git}
\bibfield{author}{\bibinfo{person}{Linus Torvalds}.}
  \bibinfo{year}{2021}\natexlab{a}.
\newblock \showarticletitle{Memory Folios commit}.
\newblock \bibinfo{journal}{\emph{github.com}} (\bibinfo{date}{Nov}
  \bibinfo{year}{2021}).
\newblock
\urldef\tempurl%
\url{https://github.com/torvalds/linux/commit/49f8275c7d92?spm=a2c65.11461447.0.0.68003853jfMprB}
\showURL{%
\tempurl}


\bibitem[Torvalds(2021b)]%
        {mmanagement_linux5.15}
\bibfield{author}{\bibinfo{person}{Linus Torvalds}.}
  \bibinfo{year}{2021}\natexlab{b}.
\newblock \showarticletitle{Memory management in Linux 5.15}.
\newblock \bibinfo{journal}{\emph{GitHub}} (\bibinfo{date}{Aug}
  \bibinfo{year}{2021}).
\newblock
\urldef\tempurl%
\url{https://github.com/torvalds/linux/blob/v5.15/Documentation/x86/x86_64/mm.rst}
\showURL{%
\tempurl}


\bibitem[W3Techs(2023)]%
        {W3Techs_OS_Server}
\bibfield{author}{\bibinfo{person}{W3Techs}.} \bibinfo{year}{2023}\natexlab{}.
\newblock \showarticletitle{Usage statistics of operating systems for
  websites}.
\newblock \bibinfo{journal}{\emph{W3Techs}} (\bibinfo{date}{Sep}
  \bibinfo{year}{2023}).
\newblock
\urldef\tempurl%
\url{https://w3techs.com/technologies/overview/operating_system}
\showURL{%
\tempurl}


\bibitem[WhatsApp(2024)]%
        {whatsapp_cves}
\bibfield{author}{\bibinfo{person}{WhatsApp}.} \bibinfo{year}{2024}\natexlab{}.
\newblock \showarticletitle{WhatsApp Security Advisories Archive}.
\newblock \bibinfo{journal}{\emph{whatsapp.com}} (\bibinfo{date}{Jun}
  \bibinfo{year}{2024}).
\newblock
\urldef\tempurl%
\url{https://www.whatsapp.com/security/advisories/archive?lang=en_US}
\showURL{%
\tempurl}


\bibitem[Whitehouse(2007)]%
        {whitehouse2007analysis}
\bibfield{author}{\bibinfo{person}{Ollie Whitehouse}.}
  \bibinfo{year}{2007}\natexlab{}.
\newblock \showarticletitle{An analysis of address space layout randomization
  on Windows Vista}.
\newblock \bibinfo{journal}{\emph{Symantec advanced threat research}}
  (\bibinfo{year}{2007}), \bibinfo{pages}{1--14}.
\newblock


\bibitem[Wilcox(2021)]%
        {folios_article}
\bibfield{author}{\bibinfo{person}{Matthew Wilcox}.}
  \bibinfo{year}{2021}\natexlab{}.
\newblock \showarticletitle{Memory Folios}.
\newblock \bibinfo{journal}{\emph{lwn.net}} (\bibinfo{date}{May}
  \bibinfo{year}{2021}).
\newblock
\urldef\tempurl%
\url{https://lwn.net/Articles/856016/}
\showURL{%
\tempurl}


\bibitem[Yu et~al\mbox{.}(2023)]%
        {DBLP:conf/uss/YuDJKF23}
\bibfield{author}{\bibinfo{person}{Jiyong Yu}, \bibinfo{person}{Aishani Dutta},
  \bibinfo{person}{Trent Jaeger}, \bibinfo{person}{David Kohlbrenner}, {and}
  \bibinfo{person}{Christopher~W. Fletcher}.} \bibinfo{year}{2023}\natexlab{}.
\newblock \showarticletitle{Synchronization Storage Channels {(S2C):}
  Timer-less Cache Side-Channel Attacks on the Apple {M1} via Hardware
  Synchronization Instructions}. In \bibinfo{booktitle}{\emph{32nd {USENIX}
  Security Symposium, {USENIX} Security 2023, Anaheim, CA, USA, August 9-11,
  2023}}, \bibfield{editor}{\bibinfo{person}{Joseph~A. Calandrino} {and}
  \bibinfo{person}{Carmela Troncoso}} (Eds.). \bibinfo{publisher}{{USENIX}
  Association}, \bibinfo{pages}{1973--1990}.
\newblock
\urldef\tempurl%
\url{https://www.usenix.org/conference/usenixsecurity23/presentation/yu-jiyong}
\showURL{%
\tempurl}


\bibitem[Zabrocki(2010)]%
        {byte-per-byte}
\bibfield{author}{\bibinfo{person}{Adam~'pi3' Zabrocki}.}
  \bibinfo{year}{2010}\natexlab{}.
\newblock \showarticletitle{Scraps of notes on remote stack overflow
  exploitation}.
\newblock \bibinfo{journal}{\emph{Phrack Magazine}} (\bibinfo{date}{Nov}
  \bibinfo{year}{2010}).
\newblock
\urldef\tempurl%
\url{http://phrack.org/issues/67/13.html\#article}
\showURL{%
\tempurl}


\bibitem[Zia et~al\mbox{.}(2023)]%
        {DBLP:journals/jaihc/ZiaAIJG23}
\bibfield{author}{\bibinfo{person}{Muhammad Zia}, \bibinfo{person}{M.~Faisal
  Amjad}, \bibinfo{person}{Zafar Iqbal}, \bibinfo{person}{Abdul~Rehman Javed},
  {and} \bibinfo{person}{Thippa~Reddy Gadekallu}.}
  \bibinfo{year}{2023}\natexlab{}.
\newblock \showarticletitle{Circumventing Google Play vetting policies: a
  stealthy cyberattack that uses incremental updates to breach privacy}.
\newblock \bibinfo{journal}{\emph{J. Ambient Intell. Humaniz. Comput.}}
  \bibinfo{volume}{14}, \bibinfo{number}{5} (\bibinfo{year}{2023}),
  \bibinfo{pages}{4785--4794}.
\newblock
\urldef\tempurl%
\url{https://doi.org/10.1007/S12652-023-04535-7}
\showDOI{\tempurl}


\end{thebibliography}

\appendix

\section{Challenges in Android ASLR Evaluation}
\label{sec:android}

Android is the most popular mobile operating system, with a market share of 72.15\% in July 2024~\cite{StatCounter_Mobile}. Unlike desktop \acp{OS}, the main threats are represented by malicious applications or malware that millions of users erroneously download from the Google Play Store~\cite{DBLP:journals/jaihc/ZiaAIJG23}. In this context, all userspace mitigations, such as stack canaries and \ac{ASLR}, are useless as the malware already has execution rights. However, userspace mitigations in Android are still relevant for all the applications that receive any external user input. In fact, applications such as messaging apps and browsers are vulnerable to memory corruption attacks~\cite{whatsapp_cves, telegram_cves}.

Evaluating the \ac{ASLR} implementation in Android is outside the scope of this work, as we focus only on desktop \acp{OS}. However, we want to introduce the main challenges in evaluating the \ac{ASLR} implementation in Android and provide some known characteristics.

One of the most important aspects of Android is that every application is forked from the \textit{Zygote} process, which is the parent process of all applications. The Zygote process loads most of the shared libraries used by the applications at boot-time to speed up the application launch time. When an application is launched, the Zygote process is forked, and the application is loaded into the virtual memory of the child process. Therefore, we expect some memory objects, such as the shared libraries, to be allocated at the same address in different applications. Additionally, due to these objects, we would need to collect samples from different reboots; an even more time-consuming task given the devices' performance.

Most recent Android phones and tablets rely on the ARMv8-A/ARMv9-A architecture, which provides a 48-bit virtual address space. Differently from desktop architectures, it is possible to choose the size of memory pages~\cite{armv8a_address_translation}, a.k.a. \textit{Translation Granule (TG)}, among three different sizes: 4KB, 16KB, and 64KB. Additionally, this customization is supported by the Android kernel~\cite{android_mmu_configurations}, which allows vendors to choose the page size according to the device's requirements. This, along with other Android customizations, potentially makes the \ac{ASLR} implementation different from one device to another. To address this issue and perform an in-depth analysis of the \ac{ASLR} implementation in Android, we would need to collect samples from different Android devices. However, we do not know if Android emulators, such as the one provided by Android Studio~\cite{android_studio}, can faithfully emulate different devices. Therefore, we cannot guarantee that samples collected from an emulator are representative of a real device. 

Another notable difference with respect to desktop \acp{OS} is that the Android default heap allocator, Scudo~\cite{android_scudo}, is not deterministic. In fact, the allocator employs the so-called \textit{chunks segregation}, a technique that groups chunks of similar sizes within the same memory page, and whenever an allocation is requested, the allocator randomly returns one of these chunks. In this way, the allocator exploits the page offset bits in the randomization process. Therefore, the analysis of heap-allocated objects in Android must consider a theoretical entropy of 47 bits. 

Finally, sampling the memory objects in Android is more challenging than in desktop \acp{OS}. In fact, Android applications are written in Java or Kotlin and run in the \textit{Android Runtime (ART)} environment. The ART environment is responsible for the memory management of the application, and it is not possible to directly access the memory objects from the application. Therefore, to overcome this limitation, we would need to write a native library in C++ to access the memory objects.

\section{Correlation Matrix Entropy}
\label{sec:corr_matrix_entropy}

\begin{figure*}
    \centering
    \includegraphics[width=\textwidth]{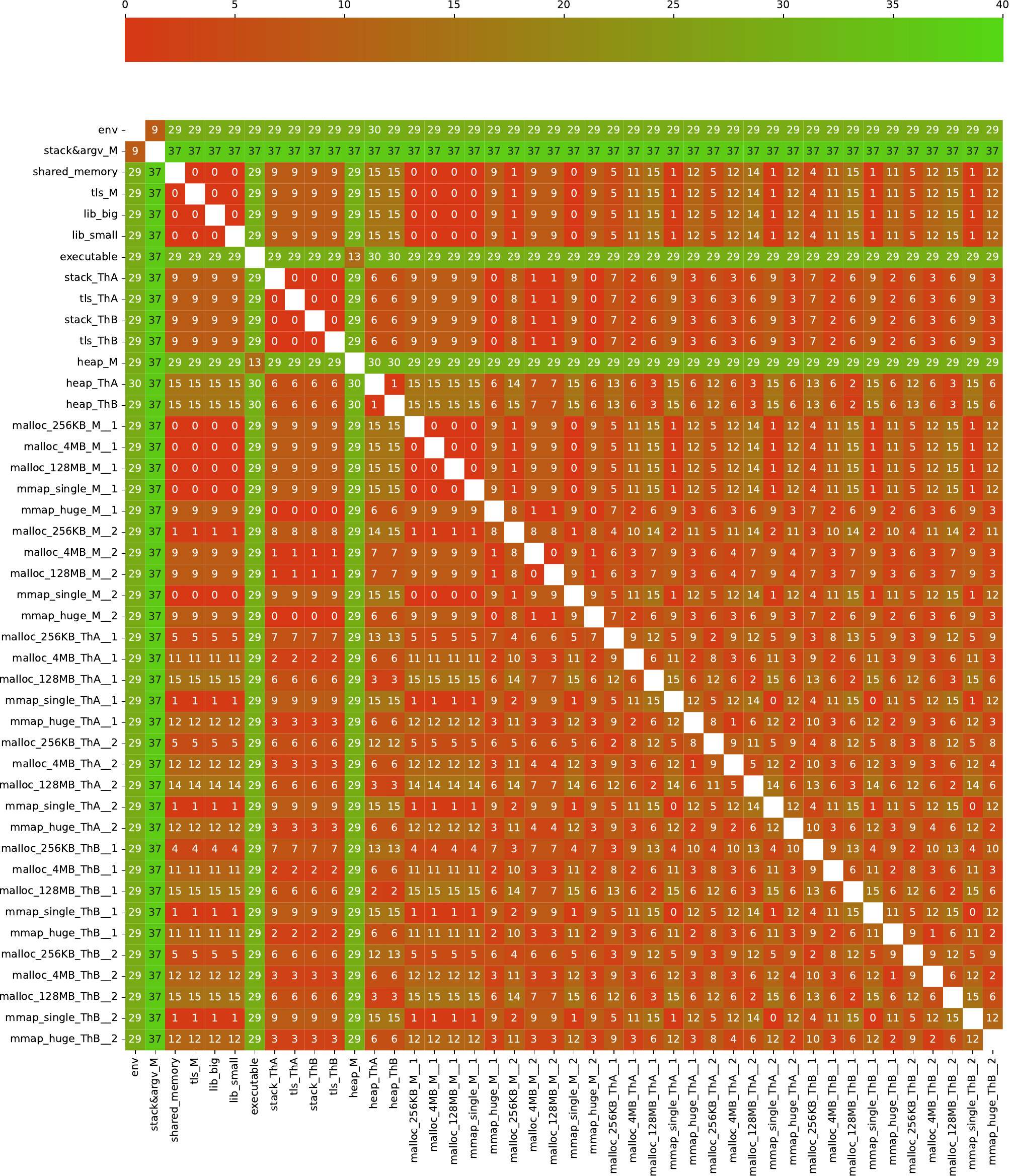}
    \caption{Correlation Entropy Linux 5.17.19}
    \label{fig:corr_lin517}
\end{figure*}

\begin{figure*}
    \centering
    \includegraphics[width=\textwidth]{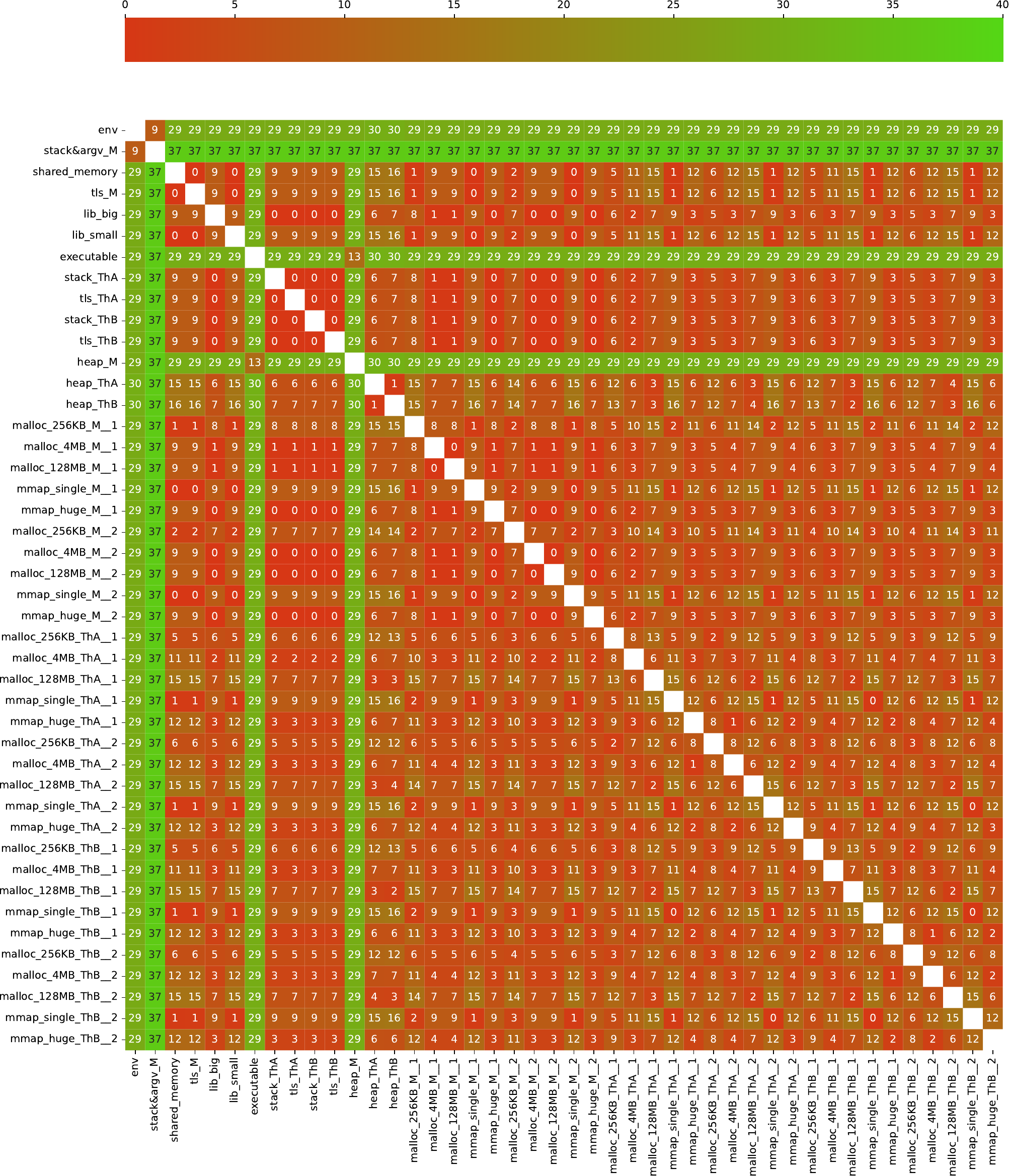}
    \caption{Correlation Entropy Linux 6.4.9}
    \label{fig:corr_lin649}
\end{figure*}

\begin{figure*}
    \centering
    \includegraphics[width=\textwidth]{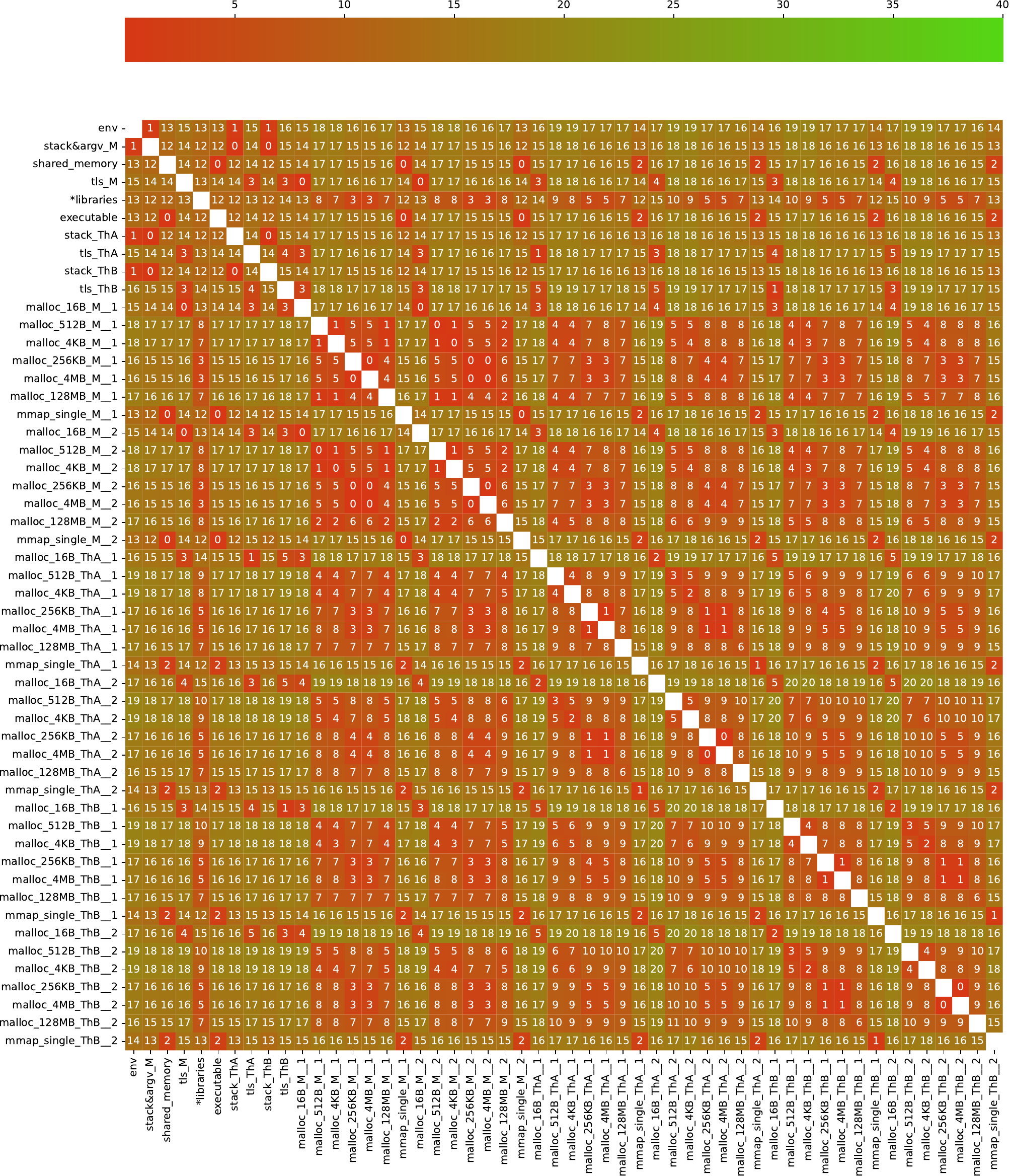}
    \caption{Correlation Entropy MacOS M1 Native}
    \label{fig:corr_macosm1_Native}
\end{figure*}

\begin{figure*}
    \centering
    \includegraphics[width=\textwidth]{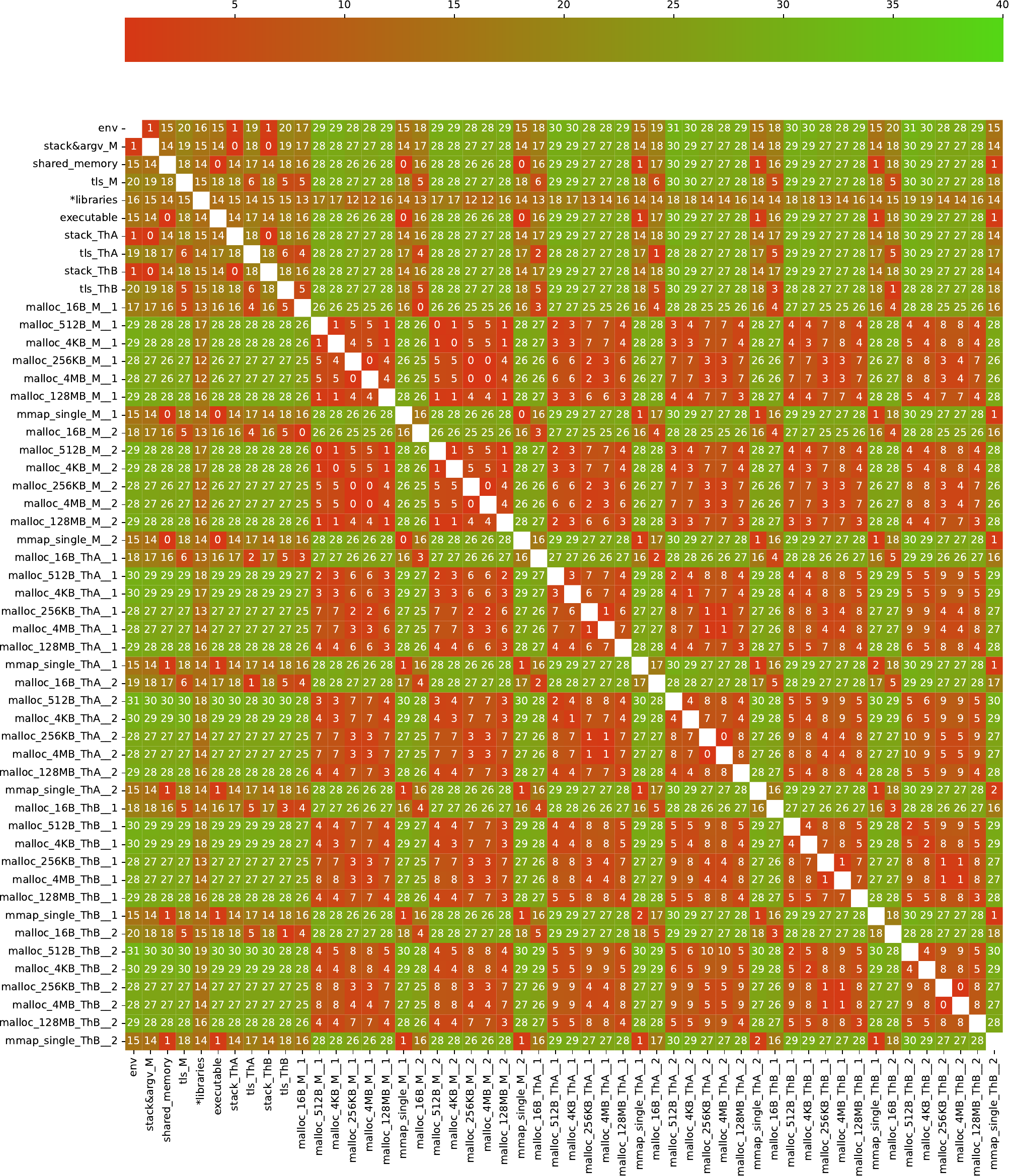}
    \caption{Correlation Entropy MacOS M1 Rosetta}
    \label{fig:correlation_macosm1_rosetta}
\end{figure*}

\begin{figure*}
    \centering
    \includegraphics[width=\textwidth]{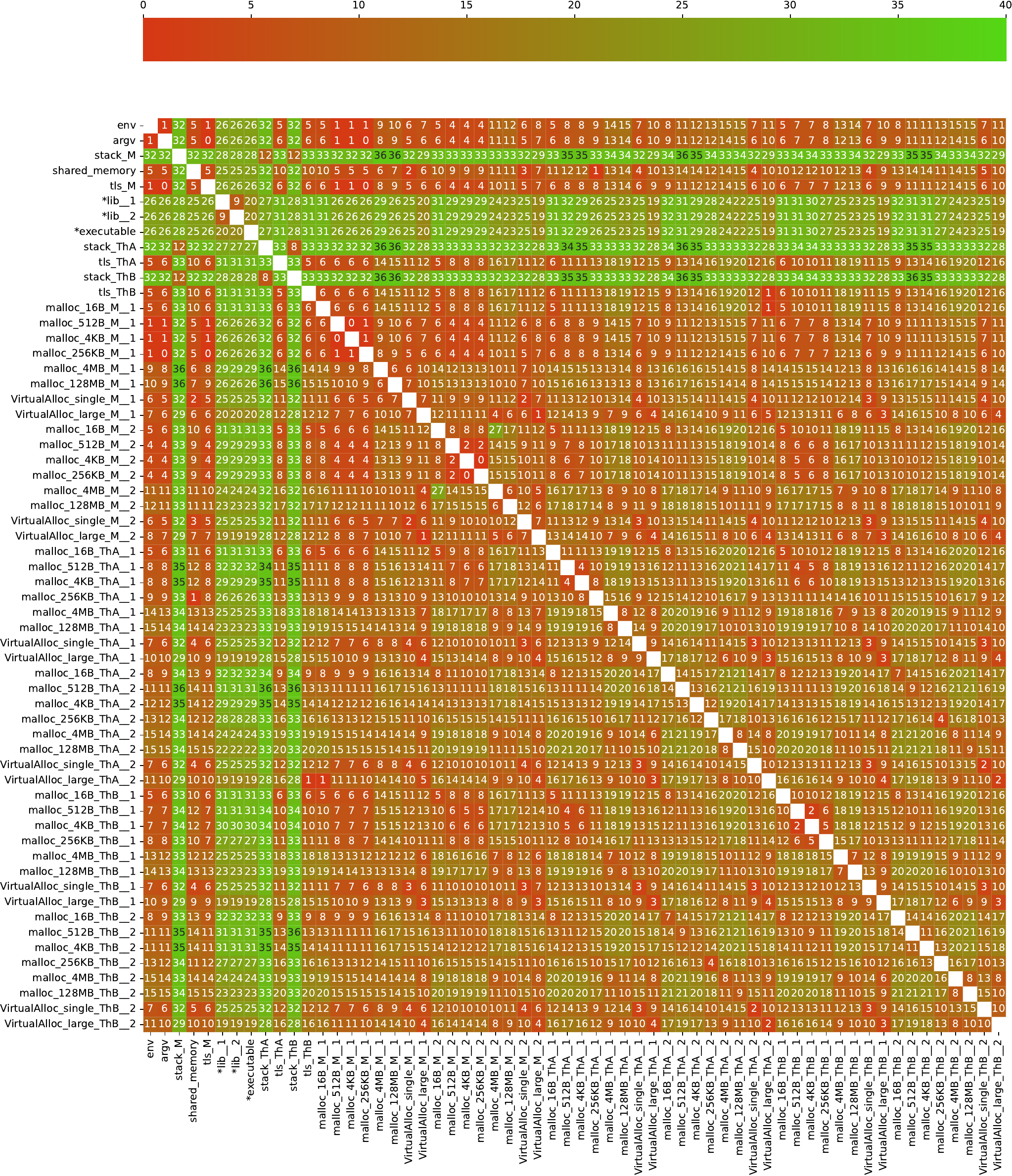}
    \caption{Correlation Entropy Windows 11}
    \label{fig:corr_matrix_win}
\end{figure*}

To support our discussion about cross entropy, we present the correlation matrix entropy for each operating system: in Figure~\ref{fig:corr_lin517} and Figure~\ref{fig:corr_lin649}, Linux with kernel 5.17.19 and 6.4.9; in Figure~\ref{fig:corr_macosm1_Native} and Figure~\ref{fig:correlation_macosm1_rosetta}, MacOS M1 native and MacOS M1 Rosetta; in Figure~\ref{fig:corr_matrix_win} Windows 11. The correlation matrix entropy is a measure of the randomness among the memory objects. The figure are aviable also at \url{https://zenodo.org/records/12968870} for a better visualization.

\end{document}